\documentclass[12pt]{article}
\usepackage{amsmath,amssymb}
\usepackage{bm}
\usepackage{color}
\usepackage{graphicx}
\allowdisplaybreaks
\newcommand{\tmop}[1]{\ensuremath{\operatorname{#1}}}

\newcommand{\bea}{\begin{eqnarray}}
\newcommand{\eea}{\end{eqnarray}}
\newcommand{\bx}{\textbf{x}}
\newcommand{\bp}{\textbf{p}}
\newcommand{\bq}{\textbf{q}}
\newcommand{\bs}{\boldsymbol{\sigma}}
\newcommand{\nn}{\nonumber}
\newcommand{\ap}{|\bp|}

\newcommand{\bn}{\textbf{n}}
\newcommand{\hEi}{\hat{E_i}}
\newcommand{\hEj}{\hat{E_j}}
\newcommand{\hq}{\hat{q}}

\begin{document}
\title{
\begin{flushright}
\ \\*[-80pt] 
\begin{minipage}{0.2\linewidth}
\normalsize
HUPD-2004 \\*[5pt]
\end{minipage}
\end{flushright}
{\Large \bf 
Time Evolution of Lepton Number Carried by Majorana Neutrinos
\\*[5pt]}}
\author{\normalsize
\centerline{
Apriadi Salim Adam$^{1}$\footnote{E-mail address: apriadi.salim.adam@lipi.go.id},
Nicholas J. Benoit$^{2}$\footnote{E-mail address: d195016@hiroshima-u.ac.jp},
Yuta Kawamura$^{2}$\footnote{E-mail address: yuta-kawamura@hiroshima-u.ac.jp},
Yamato Matsuo$^{2}$\footnote{E-mail address: ya-matsuo@hiroshima-u.ac.jp}
} \\ \normalsize
\centerline{
Takuya Morozumi$^{3,4,}$\footnote{E-mail address: morozumi@hiroshima-u.ac.jp},
Yusuke Shimizu$^{3,4}$\footnote{E-mail address: yu-shimizu@hiroshima-u.ac.jp}, 
Yuya Tokunaga$^{5}$, and 
Naoya Toyota$^{2}$\footnote{E-mail address: m191973@hiroshima-u.ac.jp}
}
\\*[10pt]
\centerline{
\begin{minipage}{\linewidth}
\begin{center}
$^1${\it \small
Research Center for Physics, Indonesian Institute of Sciences (LIPI),\\
Serpong PUSPIPTEK Area, Tangerang Selatan 15314, Indonesia} \\*[5pt]
$^2${\it \small
Graduate School of Science, Hiroshima University, 
Higashi-Hiroshima 739-8526, Japan} \\*[5pt]
$^3${\it \small
Physics Program, Graduate School of Advanced Science and Engineering, \\
Hiroshima University, Higashi-Hiroshima 739-8526, Japan} \\*[5pt]
$^4${\it \small 
Core of Research for the Energetic Universe, Hiroshima University, \\
Higashi-Hiroshima 739-8526, Japan} \\*[5pt]
$^5${\it \small
 Hakozaki, Higashi-Ku, Fukuoka,~ 812-0053,~Japan
} \\*[5pt]
\end{center}
\end{minipage}}
\\*[50pt]}
\date{
\centerline{\small \bf Abstract}
\begin{minipage}{0.9\linewidth}
\medskip
\small
We revisit the time evolution of the lepton family number for a SU(2) doublet consisting of a neutrino and a charged lepton. The lepton family number is defined through the weak basis of the SU(2) doublet, where the charged lepton mass matrix is real and diagonal.  The lepton family number carried by the neutrino is defined by the left-handed current of the neutrino family.  For this work we assume the neutrinos have Majorana mass.  This Majorana mass term is switched on at time $t=0$ and the lepton family number is evolved.  Since the operator in the flavor eigenstate is continuously connected to that of the mass eigenstate, the creation and annihilation operators for the two eigenstates are related to each other.  We compute the time evolution of all lepton family numbers by choosing a specific initial flavor eigenstate for a neutrino.  The evolution is studied for relativistic and nonrelativistic neutrinos.  The nonrelativistic region is of particular interest for the Cosmic Neutrino Background predicted from big bang models.  In that region we find the lepton family numbers are sensitive to Majorana and Dirac phases, the absolute mass, and mass hierarchy of neutrinos.
\end{minipage}
}
\begin{titlepage}
\maketitle
\thispagestyle{empty}
\end{titlepage}
\section{Introduction}
Studies of lepton number violation in the neutrino sector are classified into two categories. One is flavor oscillation \cite{Maki:1962mu} and the other is neutrinoless double $\beta$ decay \cite{Furry:1939qr}.  Where in the former category the total lepton number is conserved, in the latter it is not.  In general, neutrino oscillation has been studied among neutrinos with a SU(2) charge and a definite chirality.  Flavor oscillation studies focus on the ultra-relativistic limit. Whereas, neutrinoless double $\beta$ decay studies are focused on neutrino-antineutrino oscillation of Majorana neutrinos \cite{Majorana:1937vz,Pontecorvo:1957qd,Bahcall:1978jn,Schechter:1980gk, Xing:2013ty}.

In this paper we revisit the development of a single framework in which both phenomena, i.e. flavor oscillation and neutrino-antineutrino oscillation, are formulated together.  Previously this was developed; however, the calculation results were found to be incorrect \cite{Adam:2020abc}.  To combine both phenomena and correct the previous development we treat the phenomena from the viewpoint of lepton family number non-conservation.

We define the lepton family number for neutrinos in terms of a SU(2) doublet.  Such that the neutrino, the upper component of a SU(2) doublet, has the same family number as that of the charged lepton, the lower component.  We note that only this kind of family number is countable through the charged current weak interaction.  We also note that the lepton family number can be defined without introducing massive neutrinos, since the lepton family number is defined by the charged lepton mass eigenstates.

When the mass of the neutrinos is turned off, each of the three types of neutrinos with a specific lepton family number become massless and in an asymptotic state. Once turned on, they are no longer in an asymptotic state and each lepton family number is not conserved.  The details of non-conservation depends on the type of neutrino mass.  For Dirac neutrinos, flavor oscillation includes transitions between neutrinos with SU(2) charge to singlet right-handed neutrinos.  While for Majorana mass, neutrino-antineutrino oscillation is a transition among (anti)neutrinos in a SU(2) doublet\footnote{Similar effects have been studied for complex scalar fields as in \cite{Morozumi:2017zyz}.}.  This paper considers neutrinos to have Majorana mass.

When neutrinos are relativistic, as is in most cases, neutrino-antineutrino oscillation is suppressed \cite{Xing:2013ty}. However, it can be enhanced when the momentum carried by neutrinos is less than its rest mass.  In fact, the cosmic neutrino background (C$\nu$B) may have such a property since its typical thermal energy is predicted as $O(10^{-4})$eV for massless neutrinos.  Our present framework can be applied to relativistic and non-relativistic neutrinos.  Therefore, the C$\nu$B can be studied within our present framework.

In our work, the lepton family number carried by the neutrino is defined with a left-
handed current of the neutrino family. We study the time evolution of the lepton family
number operator for Majorana neutrino. To be definite, we introduce the mass term at $t=0$
and study the time evolution of the lepton family number for the later time.
The paper is organized as follows. In section 2, we explain how the mass term is applied. We also define the lepton family number for single flavor model and multi-flavor model, then we study their time evolution.
In section 3, the time dependent expectation value is obtained by choosing an initial
state with a definite lepton family number. The expectation value is studied numerically.  The numerical results of the time evolution of the lepton family numbers are shown. Section 4 is devoted to our concluding thoughts and
outlook.
\section{Lepton Family Number}
We start with an explanation of the measurement of lepton family number and the composition for our work.
\begin{figure}[htp]
    \centering
    \includegraphics[width=0.4\linewidth]{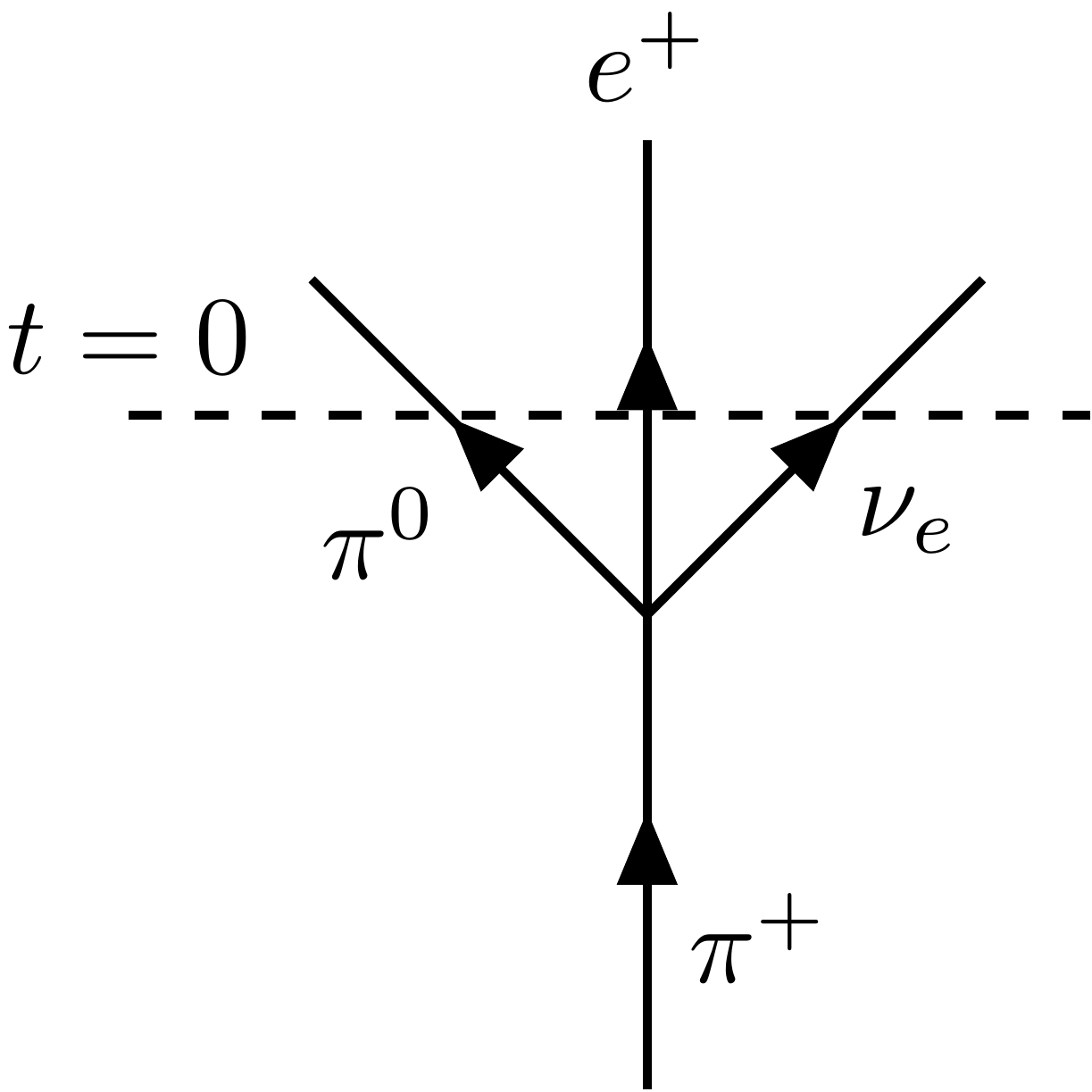}
    \caption{The production of the electron neutrino and its propagation. The neutrino acquires the Majorana mass term at $t=0$. }
    \label{Fig1}
\end{figure}
As an illustration, consider the weak decay of a charged pion; $\pi^+\rightarrow \pi^0e^+\nu_e$, as shown in Fig.\ref{Fig1}. When an electron neutrino is produced it is massless with definite flavor. Then, suppose at sometime $t=0$ the Majorana mass term is turned on.  From that time onward, the neutrino no longer has definite flavor due to mixing between the mass and flavor eigenstates.  We are interested in the time evolution of the lepton family number after the Majorana mass term is turned on.
\subsection{One-Flavor Case}\label{ssec:oneflavor}
We begin with a discussion on the one flavor case for Majorana neutrinos and show how one can define the lepton number carried by a left-handed neutrino.  This situation is described by the Lagrangian,
\bea
    \mathcal{L}=\overline{\nu_L} i \gamma^\mu \partial_\mu \nu_L -\theta(t)\frac{m}{2}\left(\overline{(\nu_L)^c} \nu_L+h.c.\right),
    \label{eq:Lag1}
\eea
where the first term is the kinetic part and the second term is the Majorana mass. The second term is controlled by a step-function to emulate what is shown in Fig.\ref{Fig1}.  This is similar to the situation previously explored for scalar fields \cite{Hotta:2014ewa}. We simplify the notation of the Lagrangian in Eq.(\ref{eq:Lag1}) by introducing the Majorana field,
\bea
    \psi=\nu_L+(\nu_L)^c,
\eea
which satisfies the Majorana condition of $\psi^c=\psi$.  This leads to the following Lagrangian,
\bea
    \mathcal{L}=\frac{1}{2}\overline{\psi} i \gamma^\mu \partial_\mu \psi -\theta(t)\frac{m}{2}\Bigl(\overline{\psi}\psi\Bigr).
    \label{eq:Lag2}
\eea
\subsubsection{Behavior in Different Time Regions}
For the region of $t<0$, the Majorana mass term is switched off.  In this region we expand the neutrino field by massless on-shell spinors.
\bea
    \nu_L(t,\bx )=\int^\prime \frac{d^3 {\bf p}}{(2\pi )^32|\bp|} \left(a(\bp) e^{-i \ap t+ i \bp \cdot \bx} u_L(\bp)+b^\dagger(\bp) e^{i \ap t-i \bp \cdot \bx}v_L(\bp) \right),
    \label{eq:massless}
\eea
where, $a(\bp)$ and $b(\bp)$ denote particle and anti-particle annihilation operators respectively.  The left-handed $\left(P_L=\frac{1-\gamma_5}{2} \right)$ massless spinors $u_L(\bp)$ and $v_L(\bp)$ satisfy the following two conditions,
\bea
    u_L(\bp)=-v_L(\bp)=\sqrt{|2\bp|}\begin{pmatrix} 0 \\ \phi_-(\bn) \end{pmatrix},
    \quad \mathbf{n} \cdot \bs \phi_{\pm}(\bn)=\pm\phi_{\pm}(\bn),
    \label{eq:masslessspinorcon}
\eea
where $\mathbf{n}=\bp/|\bp|$.  In addition, in Eq.(\ref{eq:massless}) the notation of $\int^{\prime}$ implies the zero momentum mode is excluded from the integration.  Physically this represents the absence of the zero energy state for a massless particle.  Lastly when the neutrino is massless the lepton family number is a conserved quantity,
\bea
    L(t<0) = \int d^3 \bx :\overline{\nu_L} \gamma^0 \nu_L:=\int^\prime \frac{d^3 \bp}{|2\bp|(2\pi)^3} \left(a^\dagger(\bp) a(\bp)-b^\dagger(\bp) b(\bp)\right).
    \label{eq:leptonnumbermassless}
\eea 
Since the mass term is turned on by the step function with respect to time, the translational invariance of the space direction is maintained and the space momentum is conserved. Therefore, in the region of $t>0$ the Majorana field from Eq.(\ref{eq:Lag2}) can also be expanded without the zero mode. Physically this implies a massive Majorana field of non-zero helicities,
\bea
    \psi(\bx,0_+)=\int^\prime \frac{d^3{\bf \bp}}{(2\pi)^3 2E(\bp)}\sum_{\lambda={\pm 1}}(a_M(\bp,\lambda) u(\bp, \lambda)e^{i \bp \cdot \bx}+a_M^\dagger(\bp, \lambda) v(\bp,\lambda)e^{-i \bp \cdot \bx}),
    \label{eq:massive}
\eea
where $\lambda$ denotes the helicity state and the energy of a massive particle is $E(\bp)=\sqrt{\bp^2+m^2}$.  The massive spinors of $u(\bp, \lambda)$ and $v(\bp,\lambda)$ are normalized such that they are defined with definite helicities.
\bea
    u(\bp, +1)&=&\sqrt{N(\bp)}
        \begin{pmatrix} \phi_+(\bn) \\  \frac{m}{E(\bp)+\ap} \phi_+(\bn) \end{pmatrix},
      \quad u(\bp, -1)=\sqrt{N(\bp)}
        \begin{pmatrix} \frac{m}{E(\bp)+\ap} \phi_-(\bn) \\ \phi_-(\bn) \end{pmatrix}, 
    \label{eq:massspinorcon1} \\
    v(\bp, +1)&=&\sqrt{N(\bp)}
        \begin{pmatrix} \frac{m}{E(\bp)+\ap} \phi_-(\bn) \\ -\phi_-(\bn) \end{pmatrix},
      \quad v(\bp,-1)= \sqrt{N(\bp)}
        \begin{pmatrix} -\phi_+(\bn) \\  \frac{m}{E(\bp)+\ap} \phi_+(\bn) \end{pmatrix},
    \label{eq:massspinorcon2}
\eea
where $N(\bp)=\ap+E(\bp)$ is the normalization factor.
\subsubsection{Continuity Condition}
We connect the fields defined in the regions of $t>0$ and $t<0$ and obtain a continuity condition at $t=0$.  This is achieved by integrating the equation of motion derived from Eq.(\ref{eq:Lag2}) over the infinitesimal time interval $\left[-\epsilon,\epsilon \right]$,
\bea
    \nu_L(t=-\epsilon)-P_L \psi(t=+\epsilon)=\mathcal{O}(\epsilon).
\eea
Then we take the limit of $\epsilon \rightarrow 0$ to obtain the continuity condition,
\bea
    \nu_L(t=0_-,\bx)=P_L \psi(t=0_+,\bx).
    \label{eq:con1}
\eea
We use Eq.(\ref{eq:con1}) to relate the annihilation and creation operators of Eq.(\ref{eq:massless}), i.e. $a(\bp)$, with Eq.(\ref{eq:massive}), i.e. $a_M(\bp,\lambda)$, resulting in,
\begin{gather}
    \frac{1}{\sqrt{2|\bp|}}
    \begin{pmatrix}a(\bp) \\ 
        a^{\dagger}(-\bp)
    \end{pmatrix}
    = \frac{\sqrt{N(\bp)}}{2E(\bp)}
    \begin{pmatrix}1 & \frac{i m}{E(\bp)+|\bp|}\\
        \frac{i m}{E(\bp)+|\bp|} & 1
    \end{pmatrix}
    \begin{pmatrix}a_M(\bp,-) \\ 
        a_M^{\dagger}(-\bp,-)
    \end{pmatrix}
    \label{eq:operators}
    \\
    \frac{1}{\sqrt{2|\bp|}}
    \begin{pmatrix}b(\bp) \\ 
        b^{\dagger}(-\bp)
    \end{pmatrix}
    = \frac{\sqrt{N(\bp)}}{2E(\bp)}
    \begin{pmatrix}1 & \frac{i m}{E(\bp)+|\bp|}\\
        \frac{i m}{E(\bp)+|\bp|} & 1
    \end{pmatrix}
    \begin{pmatrix}a_M(\bp,+) \\ 
        a_M^{\dagger}(-\bp,+)
    \end{pmatrix}.
    \label{eq:operators2}
\end{gather}
Details of the derivation are located in appendix \ref{sec:appendixderivation}\footnote{This derivation shows the departure from the previous work \cite{Adam:2020abc}.}.  We note the operators in Eq.(\ref{eq:operators}) and Eq.(\ref{eq:operators2}) are only valid for momenta $\left\{\bp\neq0,\bp\in A, -\bp\in \bar{A} \right\}$ as defined in the appendix \ref{sec:appendixderivation}.  The operators also obey anti-commutation relations for both massive and massless fields as shown,
\bea
    \{a(\bp),a^\dagger(\bq)\}= \{b(\bp),b^\dagger(\bq)\}=(2\pi)^3 2|\bq|\delta^3(\bp-\bq),
    \label{eq:anti-commutation} \\
    \{a_M(\bp,\lambda),a_M^\dagger(\bq,\lambda^\prime)\}=(2\pi)^3 2E(\bp)\delta^3(\bp-\bq)\delta_{\lambda\lambda'}.
    \label{eq:massanti-commutation}
\eea
Lastly, the lepton number operator for the region of $t < 0$ was derived in Eq.(\ref{eq:leptonnumbermassless}).
\subsubsection{Time Evolution}
In order to calculate the lepton number for the region of $t>0$, we must write down the time evolution of the annihilation and creation operators from Eq.(\ref{eq:operators}).  We employ that the evolution will occur in the massive region to write $a_M(\bp,\lambda)$ as $a_M(\bp,\lambda)e^{-iE(\bp)t}$ which results in,
\bea
    \frac{1}{\sqrt{2|\bp|}}
    \begin{pmatrix}a(\bp, t) \\ 
        a^{\dagger}(-\bp, t)
    \end{pmatrix}
    = \frac{\sqrt{N(\bp)}}{2E(\bp)}
    \begin{pmatrix}1 & \frac{i m}{E(\bp)+\bp}\\
        \frac{i m}{E(\bp)+\bp} & 1
    \end{pmatrix}
    \begin{pmatrix}e^{-iE(\bp)t} & 0\\
        0 & e^{+iE(\bp)t}
    \end{pmatrix}
    \begin{pmatrix}a_M(\bp,-) \\ 
        a_M^{\dagger}(-\bp,-)
    \end{pmatrix}.
    \label{eq:timeoperators}
\eea
Next, we use the result from the continuity condition in Eq.(\ref{eq:operators}) to write $a_M(\bp,\lambda)$ in terms of $a(\bp)$.
\bea
    \begin{pmatrix}a(\bp, t) \\ 
        a^{\dagger}(-\bp, t)
    \end{pmatrix}
    = \left\{ \cos{(E(\bp)t)}
    \begin{pmatrix}1 & 0\\
        0 & 1
    \end{pmatrix}
    -i\sin{(E(\bp)t)}
    \begin{pmatrix}\frac{p}{E(\bp)} & -\frac{i m}{E(\bp)}\\
        \frac{i m}{E(\bp)} & -\frac{p}{E(\bp)}
    \end{pmatrix} \right\}
    \begin{pmatrix}a(\bp) \\ 
        a^{\dagger}(-\bp)
    \end{pmatrix}.
    \label{eq:evoperators}
\eea
We note for the one-flavor case $b(\bp, t)$ and  $b^{\dagger}(-\bp, t)$ are written in the same manner as Eq.(\ref{eq:evoperators}); however, $a(\bp)$ and $a^{\dagger}(-\bp)$ are exchanged for $b(\bp)$ and $b^{\dagger}(-\bp)$ respectively.  Lastly, we rewrite the lepton number operator from Eq.(\ref{eq:leptonnumbermassless}) for the massive region of $t>0$,
\bea
    L(t) =\int^{\prime}_{\bp \in A} \left(a^\dagger(\bp,t) a(\bp,t)-b^\dagger(\bp,t) b(\bp,t)+a^\dagger(-\bp,t) a(-\bp,t)-b^\dagger(-\bp,t) b(-\bp,t)\right),
    \label{eq:massiveleptonnumber}
\eea
where we have used the notation $\int^{\prime}_{\bp \in A}=\int^{\prime}_{\bp \in A}\frac{d^3 \bp}{|2\bp|(2\pi)^3}$.  The region $A$ and $\bar{A}$ of the momentum are defined in Eq.(\ref{eq:A}) located in appendix {\ref{sec:appendixderivation}}.  The relations of Eq.(\ref{eq:evoperators}) are used to solve for the evolution of Eq.(\ref{eq:massiveleptonnumber}).  Resulting in,
\begin{equation}
 \begin{split}
    L(t)=&L(0)-2\int^{\prime}\frac{d^3 \bp}{|2\bp|(2\pi)^3} \left(\frac{m}{E(\bp)}\right)^2\sin^2{(E(\bp)t)}\left(a^\dagger(\bp)a(\bp)-b^\dagger(\bp)b(\bp)\right) \\
    &-\int^{\prime}_{\bp \in A}\frac{m\sin{(2E(\bp)t)}}{E(\bp)}\left(a(-\bp)a(\bp)+a^{\dagger}(\bp)a^{\dagger}(-\bp)-b(-\bp)b(\bp)-b^{\dagger}(\bp)b^{\dagger}(-\bp)\right) \\
    &+2\int^{\prime}_{\bp \in A} \frac{i m|\bp|}{E(\bp)^2}\sin^2{(E(\bp)t)}\left(a(-\bp)a(\bp)-a^{\dagger}(\bp)a^{\dagger}(-\bp)-b(-\bp)b(\bp)+b^{\dagger}(\bp)b^{\dagger}(-\bp)\right).
    \label{eq:singleflavorresult}
 \end{split}
\end{equation}
We can understand two consequences of the result from Eq.(\ref{eq:singleflavorresult}).  First, the lepton number is not conserved as we expect from a Majorana mass term.  Second, the the integrand is well behaved at the limit of $|\bp| \rightarrow 0_+$.
\subsection{Multi-Flavor Case}
Now, we extend the formalism of section \ref{ssec:oneflavor} to include multiple neutrino flavors.  We denote the different flavors by greek subscripts, $\nu_{\alpha L}$, in a modified form of the Lagrangian from Eq.(\ref{eq:Lag1}),
\bea
    \mathcal{L}=\overline{\nu_{\alpha L}} i\gamma^{\mu}\partial_{\mu}\nu_{\alpha L}-\theta(t) \left( \frac{m_{\alpha\beta}}{2}\overline{(\nu_{\alpha L})^c} \nu_{\beta L}+h.c.\right).
    \label{eq:multiLag1}
\eea
Again, we consider the two regions $t<0$, massless fields, and $t>0$, massive fields.  In contrast to the one-flavor case, in the region of $t>0$ the flavor eigenstates are not the same as the mass eigenstates.  We denote this difference by lowercase roman subscripts, $\nu_{iL}$.  The flavor eigenstates and mass eigenstates are related to each other by a unitary matrix $V$,
\begin{gather} \label{eq:unitaryV}
    \nu_{\alpha L}=V_{\alpha i}\nu_{iL}\\
    m_i\delta_{ij}=\left(V^T\right)_{i\alpha}m_{\alpha\beta}V_{\beta j}.
\end{gather}
We simplify the notation of Eq.(\ref{eq:multiLag1}) by introducing the Majorana field,
\bea
    \psi_i=\nu_{i L}+(\nu_{i L})^c,
\eea
which satisfies the Majorana condition of $\psi_i^c=\psi_i$.  This  leads to the Lagrangian,
\bea
    \mathcal{L}=\frac{1}{2}\overline{\psi_i} i \gamma^\mu \partial_\mu \psi_i -\theta(t)\frac{m_{i}}{2}\Bigl(\overline{\psi_i}\psi_i\Bigr).
    \label{eq:multiLag2}
\eea
We relate the two regions of Eq.(\ref{eq:multiLag1}) and Eq.(\ref{eq:multiLag2}) with the continuity condition at $t=0$,
\bea
    \nu_{L\alpha}(t=0_{-})=V_{\alpha i}P_L\psi_i(t=0_{+}).
    \label{eq:multicontinuity}
\eea
Both sides of Eq.(\ref{eq:multicontinuity}) are expanded,
\begin{gather}
    \nu_{\alpha L}(0_-,\bx )=\int^\prime \frac{d^3 {\bf p}}{(2\pi )^32|\bp|} \left(a_{\alpha}(\bp) e^{i\bp\cdot\bx}u_L(\bp)+b_{\alpha}^{\dagger}(\bp) e^{-i\bp\cdot\bx}v_L(\bp)\right),\\
    V_{\alpha i}P_L \psi_i(0_+,\bx)=V_{\alpha i} P_L \int^\prime \frac{d^3{\bf \bp}}{(2\pi)^3 2E(\bp)}\sum_{\lambda={\pm 1}}\left(a_{Mi}(\bp,\lambda)u_i(\bp, \lambda)e^{i \bp \cdot \bx}+a_{Mi}^\dagger(\bp, \lambda)v_i(\bp,\lambda)e^{-i \bp \cdot \bx}\right).
\end{gather}
We use the continuity condition of Eq.(\ref{eq:multicontinuity}) to relate the annihilation and creation operators between regions $t>0$ and $t<0$.  Similar to the single flavor case of section \ref{ssec:oneflavor}, the operators are only valid for the regions $\left\{\bp\neq0,\bp\in A, -\bp\in \bar{A} \right\}$ as defined in appendix \ref{sec:appendixderivation}.  This results in the following,
\begin{gather}
    \frac{1}{\sqrt{2|\bp|}}
    \begin{pmatrix}V_{\alpha j}^{\ast}a_{\alpha}(\bp) \\ 
        V_{\alpha j}a_{\alpha}^{\dagger}(-\bp)
    \end{pmatrix}
    = \frac{\sqrt{N_j(\bp)}}{2E_j(\bp)}
    \begin{pmatrix}1 & \frac{i m_j}{E_j(\bp)+|\bp|}\\
        \frac{i m_j}{E_j(\bp)+|\bp|} & 1
    \end{pmatrix}
    \begin{pmatrix}a_{Mj}(\bp,-) \\ 
        a_{Mj}^{\dagger}(-\bp,-)
    \end{pmatrix},
    \label{eq:multioperators} \\
    \frac{1}{\sqrt{2|\bp|}}
    \begin{pmatrix}V_{\alpha j}b_{\alpha}(\bp) \\ 
        V_{\alpha j}^{\ast}b_{\alpha}^{\dagger}(-\bp)
    \end{pmatrix}
    = \frac{\sqrt{N_j(\bp)}}{2E_j(\bp)}
    \begin{pmatrix}1 & \frac{i m_j}{E_j(\bp)+|\bp|}\\
        \frac{i m_j}{E_j(\bp)+|\bp|} & 1
    \end{pmatrix}
    \begin{pmatrix}a_{Mj}(\bp,+) \\ 
        a_{Mj}^{\dagger}(-\bp,+)
    \end{pmatrix}.
    \label{eq:multioperators2}
\end{gather}
We write the time evolution of Eq.(\ref{eq:multioperators}) in the massive region of Eq.(\ref{eq:multiLag2}) by the operators $a_{Mj}(\bp,\lambda)$.  Then, we use the relations of Eq.(\ref{eq:multioperators}) and Eq.(\ref{eq:multioperators2}) to write the result in terms of the flavor operators $a_\alpha(\bp)$ and $b_\alpha(\bp)$,
\begin{gather}
    a_\alpha(\bp, t)=V_{\alpha j}^{ }V^{\ast}_{\gamma j}\left(\cos(E_j(\bp)t)-
      \frac{i|\bp|\sin(E_j(\bp)t)}{E_j(\bp)}\right)a_\gamma(\bp)-
      V_{\alpha j}^{ }V_{\gamma j}^{ }\frac{m_j\sin(E_j(\bp)t)}{E_j(\bp)}
      a^{\dagger}_\gamma(-\bp), 
    \label{eq:gather1}\\
    a^{\dagger}_\alpha(-\bp, t)=V^{\ast}_{\alpha i}V_{\beta i}^{ }\left(\cos(E_i(\bp)t)+
      \frac{i|\bp|\sin(E_i(\bp)t)}{E_i(\bp)}\right)a^{\dagger}_\beta(-\bp)+
      V^{\ast}_{\alpha i}V^{\ast}_{\beta i}\frac{m_i\sin(E_i(\bp)t)}{E_i(\bp)}
      a_\beta(\bp),
    \label{eq:gather2}\\
    b_\alpha(\bp, t)=V^{\ast}_{\alpha j}V_{\gamma j}^{ }\left(\cos(E_j(\bp)t)-
      \frac{i|\bp|\sin(E_j(\bp)t)}{E_j(\bp)}\right)b_\gamma(\bp)-
      V^{\ast}_{\alpha j}V^{\ast}_{\gamma j}\frac{m_j\sin(E_j(\bp)t)}{E_j(\bp)}
      b^{\dagger}_\gamma(-\bp),
    \label{eq:gather3}\\
    b^{\dagger}_\alpha(-\bp, t)=V_{\alpha i}^{ }V^{\ast}_{\beta i}\left(\cos(E_i(\bp)t)+
      \frac{i|\bp|\sin(E_i(\bp)t)}{E_i(\bp)}\right)b^{\dagger}_\beta(-\bp)+
      V_{\alpha i}^{ }V_{\beta i}^{ }\frac{m_i\sin(E_i(\bp)t)}{E_i(\bp)}
      b_\beta(\bp).
    \label{eq:gather4}
\end{gather}
Where the indices of $\beta$, $\gamma$, $i$, and $j$ are summed over based on the greek letters for flavor eigenstates and the roman letters for mass eigenstates.  The values of Eqs. (\ref{eq:gather1}), (\ref{eq:gather2}), (\ref{eq:gather3}) and (\ref{eq:gather4}) enable us to solve for the multi-flavor version of Eq.(\ref{eq:massiveleptonnumber}),
\begin{equation}
 \begin{split}
    L_{\alpha}(t)=& \int^{\prime}_{\bp \in A} \left[-\frac{m_im_j\sin(E_i(\bp)t)\sin(E_j(\bp)t)}{E_i(\bp)E_j(\bp)}
      \left\{V^{\ast}_{\alpha i}V^{\ast}_{\beta i}V_{\alpha j}^{}V_{\gamma j}^{}
        \left(a_{\gamma}^{\dagger}(\bp)a_{\beta}(\bp)+a_{\gamma}^{\dagger}(-\bp)a_{\beta}(-\bp)\right)\right.\right.\\
    &\phantom{XXXXXXXXXXXXXXX}
      \left.-V_{\alpha i}^{ }V_{\beta i}^{ }V^{\ast}_{\alpha j}V^{\ast}_{\gamma j}
        \left(b_{\gamma}^{\dagger}(\bp)b_{\beta}(\bp)+b_{\gamma}^{\dagger}(-\bp)b_{\beta}(-\bp)\right)\right\}\\
    &\phantom{XXX}
      +\left(\cos(E_i(\bp)t)\cos(E_j(\bp)t)+\frac{|\bp|^2}{E_i(\bp)E_j(\bp)}\sin(E_i(\bp)t)\sin(E_j(\bp)t)\right.\\
    &\phantom{XXXXXXXXx}
      \left.+\frac{i|\bp|}{E_i(\bp)}\sin(E_i(\bp)t)\cos(E_j(\bp)t)
        -\frac{i|\bp|}{E_j(\bp)}\cos(E_i(\bp)t)\sin(E_j(\bp)t)\right)\\
    &\phantom{XXXXXX}
      \times\left\{V^{\ast}_{\alpha i}V_{\beta i}^{ }V_{\alpha j}^{ }V^{\ast}_{\gamma j}
        \left(a_{\beta}^{\dagger}(\bp)a_{\gamma}(\bp)+a^{\dagger}_{\beta}(-\bp)a_{\gamma}(-\bp)\right)\right.\\
    &\phantom{XXXXXXXXXXXXXXX}
      \left.-V_{\alpha i}^{ }V^{\ast}_{\beta i}V^{\ast}_{\alpha j}V_{\gamma j}^{ }
        \left(b^{\dagger}_{\beta}(\bp)b_{\gamma}(\bp)+b^{\dagger}_{\beta}(-\bp)b_{\gamma}(-\bp)\right)\right\}\\
    &\phantom{XXX}
      -\left(\frac{m_j\sin(E_j(\bp)t)}{E_j(\bp)}\left(\cos(E_i(\bp)t)
        +\frac{i|\bp|}{E_i(\bp)}\sin(E_i(\bp)t)\right)\right)\\
    &\phantom{XXXXXX}
      \times\left\{V^{\ast}_{\alpha i}V_{\beta i}^{}V_{\alpha j}^{}V_{\gamma j}^{}
        \left(a^{\dagger}_{\beta}(\bp)a^{\dagger}_{\gamma}(-\bp)
          -a^{\dagger}_{\beta}(-\bp)a^{\dagger}_{\gamma}(\bp)\right) \right.\\
    &\phantom{XXXXXXXXXXXXXXX}
      \left.-V_{\alpha i}^{}V^{\ast}_{\beta i}V^{\ast}_{\alpha j}V^{\ast}_{\gamma j}
        \left(b^{\dagger}_{\beta}(\bp)b^{\dagger}_{\gamma}(-\bp)
          -b^{\dagger}_{\beta}(-\bp)b^{\dagger}_{\gamma}(\bp)\right) \right\} \\
    &\phantom{XXX}
      +\left(\frac{m_i\sin(E_i(\bp)t)}{E_i(\bp)}\left(\cos(E_j(\bp)t)
        -\frac{i|\bp|}{E_j(\bp)}\sin(E_j(\bp)t)\right)\right)\\
    &\phantom{XXXXXX}
      \times\left\{V^{\ast}_{\alpha i}V^{\ast}_{\beta i}V_{\alpha j}^{ }V^{\ast}_{\gamma j}
        \left(a_{\beta}(-\bp)a_{\gamma}(\bp)-a_{\beta}(\bp)a_{\gamma}(-\bp)\right) \right.\\
    &\phantom{XXXXXXXXXXXXXXX}
      \Biggl.\left.-V_{\alpha i}^{ }V_{\beta i}^{ }V^{\ast}_{\alpha j}V_{\gamma j}^{ }
        \left(b_{\beta}(-\bp)b_{\gamma}(\bp)-b_{\beta}(\bp)b_{\gamma}(-\bp)\right)\right\}\Biggr].
  \label{eq:multimassiveleptonnumber}
 \end{split}
\end{equation}
We take the summation of $\sum_{\alpha}L_{\alpha}(t)$ to study if the lepton family operator of Eq.(\ref{eq:multimassiveleptonnumber}) is a conserved value,
\begin{equation}
 \begin{split}
    L(t)=&\sum_\alpha L_\alpha(0)-\int^{\prime}\frac{d^3 \bp}{(2\pi)^3|2\bp|}\frac{2m^2_i\sin^2(E_i(\bp)t)}{E^2_i(\bp)}
      \Bigl(V_{\beta i}^{ }V^{\ast}_{\gamma i}a^{\dagger}_{\beta}(\bp)a_{\gamma}(\bp)
        -V^{\ast}_{\beta i}V_{\gamma i}^{ }b_{\beta}^\dagger(\bp)b_{\gamma}(\bp)\Bigr) \\
    &+\int^{\prime}_{\bp \in A}\frac{m_i\sin(2E_i(\bp)t)}{E_i(\bp)}
      \Bigl(V_{\beta i}^{ }V_{\gamma i}^{ }a^{\dagger}_{\beta}(\bp)a^{\dagger}_{\gamma}(-\bp)
        -V_{\beta i}^{\ast}V_{\gamma i}^{\ast}b^{\dagger}_\beta(\bp)b^{\dagger}_{\gamma}(-\bp) \Bigr. \\
    &\Bigl. \phantom{\frac{m\sin{(2E_i(\bp)t)}}{E_i(\bp)}(V_{\beta i}V_{\gamma i}a^{\dagger}_{\beta}(\bp)a^{\dagger}_{\gamma}(-\bp)}
      +V_{\beta i}^{\ast}V_{\gamma i}^{\ast}a_{\beta}(-\bp)a_{\gamma}(\bp)
        -V_{\beta i}^{ }V_{\gamma i}^{ }b_{\beta}(-\bp)b_{\gamma}(\bp)\Bigr) \\
    &+2i\int^{\prime}_{\bp \in A} \frac{m_i|\bp|\sin^2{(E_i(\bp)t)}}{E_i^2(\bp)}
      \Bigl(V_{\beta i}^{ }V_{\gamma i}^{ }a_{\beta}^{\dagger}(\bp)a_{\gamma}^{\dagger}(-\bp)
        -V_{\beta i}^{\ast}V_{\gamma i}^{\ast}b^{\dagger}_{\beta}(\bp)b^{\dagger}_{\gamma}(-\bp)\Bigr. \\
    &\Bigl.\phantom{\frac{m\sin(2E_i(\bp)t)}{E_i(\bp)}(V_{\beta i}V_{\gamma i}a^{\dagger}_{\beta}(\bp)a^{\dagger}_{\gamma}(-\bp)}
      -V_{\beta i}^{\ast}V_{\gamma i}^{\ast}a_{\beta}(-\bp)a_{\gamma}(\bp)
        +V_{\beta i}^{ }V_{\gamma i}^{ }b_{\beta}(-\bp)b_{\gamma}(\bp)\Bigr)
    \label{eq:multiflavorresult}
 \end{split}
\end{equation}
Where we have used the notation $\int^{\prime}_{\bp \in A}=\int^{\prime}_{\bp \in A}\frac{d^3 \bp}{(2\pi)^3|2\bp|}$ and the unitary property of the matrix $\sum_{\alpha}V^{\ast}_{\alpha i}V_{\alpha j}=\delta_{ij}$.  Similar to the single-flavor result of Eq.(\ref{eq:singleflavorresult}) we find that the total lepton number $L(t)$ is not conserved in Eq.(\ref{eq:multiflavorresult}).  Lepton number non-conservation for $t>0$ is allowable in our framework, because we are considering a Majorana mass term in the Lagrangians of Eq.(\ref{eq:multiLag1}) and (\ref{eq:multiLag2}).
\subsubsection{Evolution from a Single Flavor Eigenstate}
Now we consider the evolution starting from a known flavor eigenstate denoted as,
\bea
    \left| \bq,\sigma \rangle\right. = \frac{a_{\sigma}^{\dagger}(\bq)\left|0\rangle \right.}
      {\sqrt{(2\pi)^3\delta^{(3)}(0)2|\bq|}}.
  \label{eq:flavoreigenstate}
\eea
We sandwich Eq.(\ref{eq:flavoreigenstate}) around $L_{\alpha}(t)$ of Eq.(\ref{eq:multimassiveleptonnumber}) to produce a type of expectation value,
\bea
    \left.\langle\bq,\sigma\right|L_{\alpha}(t)\left|\bq,\sigma\rangle\right.
      =\frac{\left.\langle 0|a_{\sigma}(\bq) L_{\alpha}(t)a_{\sigma}^{\dagger}(\bq)
      |0\rangle\right.}{(2\pi)^3\delta^{(3)}(0)2|\bq|}.
  \label{eq:expectationvalue}
\eea
We note terms inside $\left.\langle0|a_{\sigma}(\bq) L_{\alpha}(t)a_{\sigma}^{\dagger}(\bq)|0 \rangle\right.$ evaluate either  zero or non-zero.
This is due to anti-commutation relations of operators in Eq.(\ref{eq:multioperators}) and Eq.(\ref{eq:multioperators2}) with $a_{\sigma}(\bq)$.
The calculation of Eq.(\ref{eq:expectationvalue}) then results in,
\begin{multline}
    \left.\langle\bq,\sigma\right|L_{\alpha}(t)\left|\bq,\sigma\rangle\right. =
      \sum_i \left|V_{\alpha i}\right|^2\left|V_{\sigma i}\right|^2 \frac{|\bq|^2+m_i^2\cos{(2E_i(\bq)t)}}{E_i^2(\bq)} \\
    +\sum_{\{i,j\}}\tmop{Re}\left[V_{\alpha i}^{\ast}V_{\sigma i}^{ }V_{\alpha j}^{ }V_{\sigma j}^{\ast}\right]
      \left[\cos((E_i(\bq)-E_j(\bq))t) \left(1+\frac{|\bq|^2-m_im_j
      \tmop{Re}\left[\frac{V_{\sigma i}^{\ast}V_{\sigma j}^{ }}
      {V_{\sigma i}^{ }V^{\ast}_{\sigma j}}\right]}{E_iE_j}\right) \right. \\
    \shoveright{+\left.\cos{((E_i(\bq)+E_j(\bq))t)} \left(1-\frac{|\bq|^2-m_im_j
        \tmop{Re}\left[\frac{V_{\sigma i}^{\ast}V_{\sigma j}^{ }}
        {V_{\sigma i}^{ }V^{\ast}_{\sigma j}}\right]}{E_iE_j}\right) \right]} \\
    -\tmop{Im}\left[V_{\alpha 1}^{\ast}V_{\sigma 1}^{ }V_{\alpha 2}^{ }V_{\sigma 2}^{\ast}\right]
      \sum_{\{i,j\}}\Biggl[\left(\frac{|\bq|}{E_i(\bq)}-
        \frac{|\bq|}{E_j(\bq)}\right)\sin{((E_i(\bq)+E_j(\bq))t)}\Biggr.\\
    \shoveright{+\left(\frac{|\bq|}{E_i(\bq)}+\frac{|\bq|}{E_j(\bq)}\right)\sin{((E_i(\bq)-E_j(\bq))t)}}\\
    -\frac{m_im_j}{E_i(\bq)E_j(\bq)}\tmop{Im}\left[\frac{V_{\sigma i}^{\ast}V_{\sigma j}^{}}
        {V_{\sigma i}^{}V_{\sigma j}^{\ast}}\right]\Bigl(\cos{((E_i(\bq)-E_j(\bq))t)}-\cos{((E_i(\bq)+E_j(\bq))t)}\Bigr) \Biggr] ,
  \label{eq:expectationresult}
\end{multline}
where we identify the Greek index of $\sigma$ to hold the 3 flavor content of the Standard Model of $e$, $\mu$, and $\tau$.
Consequently, from Eq.(\ref{eq:unitaryV}) the roman indices of $i$ and $j$ take on the mass eigenstates of 1, 2, and 3,
\begin{gather}
    \sum_{\sigma}=\sum_{\sigma=e,\mu}^{\tau}, \phantom{XXX} \sum_{i}=\sum_{i=1}^{3}, \\
    \sum_{\{i,j\}}=\sum_{\{i,j\}=\{1,2\};\{2,3\};\{3,1\}}.
\end{gather}
If we consider the case where the unitary matrix $V$ to be the $3\times3$ PMNS matrix the imaginary term has the property of,
\bea
    \tmop{Im}\left[V^{\ast}_{\alpha 1}V_{\sigma 1}^{ }V_{\alpha 2}^{ }V^{\ast}_{\sigma 2}\right] =
      \tmop{Im}\left[V^{\ast}_{\alpha 2}V_{\sigma 2}^{ }V_{\alpha 3}^{ }V^{\ast}_{\sigma 3}\right] =
       \tmop{Im}\left[V^{\ast}_{\alpha 3}V_{\sigma 3}^{ }V_{\alpha 1}^{ }V^{\ast}_{\sigma 1}\right] .
\eea
We evaluate Eq.(\ref{eq:expectationresult}) for three different conditions.  First, similar to the single-flavor case of Eq.(\ref{eq:singleflavorresult}), we take the summation over the index of $\alpha$.
\bea
    \sum_{\alpha}\left.\langle\bq,\sigma\right|L_{\alpha}(t)\left|\bq,\sigma\rangle\right.=
      \sum_i |V_{\sigma i}^{ }|^2\left(\frac{|\bq|^2+m_i^2\cos{2(E_i(\bq)t)}}{E^2_i(\bq)}\right)
  \label{eq:expectationsum}
\eea
We find, after the summation over $\alpha$, the expectation value of the lepton number is not conserved.  However, Eq.(\ref{eq:expectationsum}) does place limits on the lepton number expectation value,
\bea
    -1 \leq \sum_i|V_{\sigma i}^{ }|^2\frac{|\bq|^2-m_i^2}{|\bq|^2+m_i^2} \leq
      \sum_{\alpha}\left.\langle\bq,\sigma\right|L_{\alpha}(t)\left|\bq,\sigma\rangle\right. \leq 1
\eea
The second and third conditions are the ultra-relativistic limit of $\bq^2\gg m_im_j$ and zero momentum limit $|\bq| \rightarrow 0_+$ respectively.
\begin{multline}
    \lim_{|\bq|\gg m_im_j}\left.\langle\bq,\sigma\right|L_{\alpha}(t)\left|\bq,\sigma\rangle\right.
      \rightarrow \delta_{\sigma\alpha}
      -4\sum_{\{i,j\}}\tmop{Re}\left[V_{\alpha i}^{\ast}V_{\sigma i}^{}V_{\alpha j}^{}V_{\sigma j}^{\ast}\right]
        \sin^2{\left(\frac{\Delta m^2_{ij}t}{4|\bq|}\right)}\\
    -2\tmop{Im}\left[V_{\alpha 1}^{\ast}V_{\sigma 1}^{}V_{\alpha 2}^{}V_{\sigma 2}^{\ast}\right]
      \sum_{\{i,j\}}\sin\left(\frac{\Delta m_{ij}^2t}{2|\bq|}\right)
  \label{eq:ultralimit}
\end{multline}
Which, is equivalent to the standard probability transition for neutrino oscillations written as $P_{\sigma\rightarrow\alpha}(t)\ge0$.
\begin{multline}
    \lim_{|\bq|\rightarrow0_+}\left.\langle\bq,\sigma\right|L_{\alpha}(t)
      \left|\bq,\sigma\rangle\right.\rightarrow
      \sum_{i}|V_{\alpha i}^{}|^2|V_{\sigma i}^{}|^2\cos{(2m_it)}\\
    +\sum_{\{i,j\}}\tmop{Re}\left[V_{\alpha i}^{\ast}V_{\sigma i}^{}
      V_{\alpha j}^{}V_{\sigma j}^{\ast} \right]
      \left[\cos{((m_i-m_j)t)}\left(1-\tmop{Re}\left[\frac{V_{\sigma i}^{\ast}V_{\sigma j}^{}}
      {V_{\sigma i}^{}V_{\sigma j}^{\ast}}\right]\right)\right. \\
    \shoveright{\left. +\cos{((m_i+m_j)t)}\left(1+\tmop{Re}\left[\frac{V_{\sigma i}^{\ast}V_{\sigma j}^{}}
      {V_{\sigma i}^{}V_{\sigma j}^{\ast}}\right]\right)\right]} \\
    +\tmop{Im}\left[V_{\alpha 1}^{\ast}V_{\sigma 1}^{}V_{\alpha 2}^{ }V_{\sigma 2}^{\ast}\right]
      \sum_{\{ i,j\}}\left[\cos{((m_i-m_j)t)}-\cos{((m_i+m_j)t)}\right]
      \tmop{Im}\left[\frac{V_{\sigma i}^{\ast}V_{\sigma j}^{}}
      {V_{\sigma i}^{}V_{\sigma j}^{\ast}} \right].
\label{eq:zerolimit}
\end{multline}
\section{Numerical Calculations}
From Eq.(\ref{eq:unitaryV}) we identify the unitary matrix $V$ as the PMNS matrix with two additional Majorana phases $\alpha_{21}$ and $\alpha_{31}$,
\bea
    V_{\sigma i}=
    \begin{pmatrix}
        c_{12}c_{13} & s_{12}c_{13} & s_{13}e^{-i\delta}\\
        -s_{12}c_{23}-c_{12}s_{23}s_{13}e^{i\delta} & c_{12}c_{23}-s_{12}s_{23}s_{13}e^{i\delta} & s_{23}c_{13} \\
        s_{12}s_{23}-c_{12}c_{23}s_{13}e^{i\delta} & -c_{12}s_{23}-s_{12}c_{23}s_{13}e^{i\delta} & c_{23}c_{13}
    \end{pmatrix}
    \begin{pmatrix}
     1 & 0 & 0 \\
     0 & e^{i\frac{\alpha_{21}}{2}} & 0 \\
     0 & 0 & e^{i\frac{\alpha_{31}}{2}}
    \end{pmatrix},
  \label{eq:PMNSmatrix}
\eea
where we use the notation adopted from \cite{Zyla:2020zbs} for the mixing angles $c_{13}=\cos{\theta_{13}}$, $s_{12}=\sin{\theta_{12}}$, etc., and $\delta$ as the CP violating phase \cite{Kobayashi:1973fv}.
The values for the mixing angles and the CP violating phase are experimentally constrained.  We choose to use the best fit values from $\nu$fit collaboration data 5.0 \cite{Esteban:2020cvm} unless otherwise stated.  Thus in Eq.(\ref{eq:PMNSmatrix}), we are left with the freedom to choose values for the Majorana phases $\alpha_{21}$ and $\alpha_{31}$ \cite{Doi:1980yb}.

Next, precise values of the neutrino mass eigenstates are not experimentally known.  However, the mass squared differences ($\Delta m^2_{ij}$) are experimentally constrained \cite{Esteban:2016qun} and an upper limit for the  summation of the mass eigenstates is placed from cosmology observations \cite{Aghanim:2018eyx}.  Again, we use the best fit values from $\nu$fit collaboration data 5.0 \cite{Esteban:2020cvm} unless otherwise stated.  Thus for Eq.(\ref{eq:expectationresult}) we are left with choices on the hierarchy of masses, Normal ($m_3\gg m_2>m_1$) or Inverted ($m_2>m_1\gg m_3$), and the value of the lightest mass.
\subsection{Setup}
We expand the sum of the contribution from the three mass eigenstates in
Eq.(\ref{eq:expectationresult}) as follows,
\begin{multline}
    \langle\bq,\sigma|L_{\alpha}(t)|\bq,\sigma\rangle = |V_{\alpha 1}^{\ast}|^2|   
       V_{\sigma 1}^{} |^2 \left(1-\frac{2m_1^2 \sin^2(E_1(\bq)t)}{E_1^2(\bq)}  \right) \\
    +|V_{\alpha 2}^{\ast}|^2|
       V_{\sigma 2}^{} |^2\left(1-\frac{2m_2^2  \sin^2(E_2(\bq)t)}{E_2^2(\bq)} \right) 
       +|V_{\alpha 3}^{\ast}|^2 |V_{\sigma 3}^{}|^2 
       \left(1-\frac{2m_3^2  \sin^2(E_3(\bq)t)}{E_3^2(\bq)} \right) \\
    +\tmop{Re} [V_{\alpha 1}^{\ast}V_{\sigma 1}^{} V_{\alpha 2}^{} V_{\sigma 2}^{\ast}]
       \left\{ \left( 1+\frac{| \bq |^2-m_1 m_2
       \tmop{Re}\left[ \frac{V_{\sigma 1}^{\ast} V_{\sigma 2}^{}}{V_{\sigma 1}^{}
       V^{\ast}_{\sigma 2}} \right] }{E_1(\bq)E_2(\bq)}  \right) \cos (E_1(\bq) -
       E_2(\bq)) t  \right.\\
    +\left. \left(1 - \frac{|\bq |^2-m_1 m_2 \tmop{Re}\left[
       \frac{V_{\sigma 1}^{\ast} V_{\sigma 2}^{}}{V_{\sigma 1}^{}
       V^{\ast}_{\sigma 2}} \right] }{E_1(\bq)E_2(\bq)}  \right) \cos (E_1(\bq) +
       E_2(\bq)) t \right \} \\
    +\tmop{Re} [V_{\alpha 2}^{\ast}V_{\sigma 2}^{} V_{\alpha 3}^{} V_{\sigma 3}^{\ast}]
       \left\{ \left( 1 + \frac{| \bq |^2-m_2 m_3 \tmop{Re}\left[
       \frac{V_{\sigma 2}^{\ast} V_{\sigma 3}^{}}{V_{\sigma 2}^{}
       V^{\ast}_{\sigma 3}} \right] }{E_2(\bq)E_3(\bq)}  \right) \cos (E_2(\bq) -
       E_3(\bq)) t  \right. \\
    +\left. \left(1 - \frac{|\bq |^2-m_2 m_3 \tmop{Re}\left[\frac{V_{\sigma 2}^{\ast}
       V_{\sigma 3}^{}}{V_{\sigma 2}^{}V^{\ast}_{\sigma 3}}
       \right] }{E_2(\bq)E_3(\bq)}  \right) \cos (E_2(\bq) +
       E_3(\bq)) t \right \} \\
    +\tmop{Re}[V_{\alpha 3}^{\ast}V_{\sigma 3}^{} V_{\alpha 1}^{} V_{\sigma 1}^{\ast}]
       \left\{ \left( 1 + \frac{| \bq |^2-m_3 m_1 \tmop{Re}\left[
       \frac{V_{\sigma 3}^{\ast} V_{\sigma 1}^{}}{V_{\sigma 3}^{}
       V^{\ast}_{\sigma 1}} \right] }{E_3(\bq)E_1(\bq)}  \right) \cos (E_3(\bq) -
       E_1(\bq)) t  \right. \\
    +\left. \left(1 - \frac{|\bq |^2-m_3 m_1 \tmop{Re}\left[
       \frac{V_{\sigma 3}^{\ast} V_{\sigma 1}}{V_{\sigma 3}
       V^{\ast}_{\sigma 1}} \right] }{E_3(\bq)E_1(\bq)}  \right) \cos (E_3(\bq) +
       E_1(\bq)) t \right \} \\
    -\tmop{Im}[V_{\alpha 1}^{}V_{\sigma 1}^{\ast} V^{\ast}_{\alpha 2} V^{}_{\sigma 2}]
       \Biggl\{ - \tmop{Im} \left[
       \frac{V_{\sigma 1}^{\ast} V_{\sigma 2}}{V_{\sigma 1}^{} V^{\ast}_{\sigma 2}}
       \right] \frac{m_1 m_2 (\cos (E_{1}(\bq) - E_{2}(\bq)) t - \cos (E_{1}(\bq) +
       E_{2}(\bq)) t)}{E_{1}(\bq) E_{2}(\bq)} \\
    -\tmop{Im} \left[
       \frac{V_{\sigma 2}^{\ast} V_{\sigma 3}^{}}{V_{\sigma 2}^{} V^{\ast}_{\sigma 3}}
       \right] \frac{m_2 m_3 (\cos (E_{2}(\bq) - E_{3}(\bq)) t - \cos (E_{2}(\bq) +
       E_{3}(\bq)) t)}{E_{2}(\bq) E_{3}(\bq)} \\
    -\tmop{Im} \left[
       \frac{V_{\sigma 3}^{\ast} V_{\sigma 1}^{}}{V_{\sigma 3}^{} V^{\ast}_{\sigma 1}}
       \right] \frac{m_3 m_1 (\cos (E_{3}(\bq) - E_{1}(\bq)) t - \cos (E_{3}(\bq) +
       E_{1}(\bq)) t)}{E_{3}(\bq) E_{1}(\bq) } \\
    +\left(\frac{|\bq |}{E_1 (\bq)} -\frac{|\bq |}{E_2(\bq)} \right)
       \sin (E_{1}(\bq) +E_{2}(\bq)) t + 
       \left( \frac{|\bq |}{E_1 (\bq)} +\frac{|\bq |}{E_2(\bq)} \right)
       \sin(E_{1}(\bq)-E_{2}(\bq)) t ) \\
    +\left(\frac{|\bq |}{E_2 (\bq)} -\frac{|\bq |}{E_3(\bq)} \right)
       \sin (E_{2}(\bq)+E_{3}(\bq)) t + 
       \left( \frac{|\bq |}{E_2(\bq)} +\frac{|\bq |}{E_3(\bq)} \right)
       \sin(E_{2}(\bq)-E_{3}(\bq)) t ) \\
    +\left(\frac{|\bq |}{E_3 (\bq)} -\frac{|\bq |}{E_1(\bq)} \right)
       \sin (E_{3}(\bq)+E_{1}(\bq)) t + 
       \left( \frac{|\bq |}{E_3(\bq)} +\frac{|\bq |}{E_1(\bq)} \right)
       \sin(E_{3}(\bq)-E_{1}(\bq)) t ) \Biggl\}. \\
  \label{eq:Lsummass}
\end{multline}
We note the combinations of the PMNS matrix which have the form 
$V_{\alpha i}^{\ast}V_{\sigma i}^{} V_{\alpha j}^{} V_{\sigma j}^{\ast}$
are independent of Majorana phases. While the other combinations of the form
$ \frac{V_{\sigma i}^{\ast} V_{\sigma j}}{V_{\sigma i}^{} V^{\ast}_{\sigma j}}$
depend on the Majorana phases.  Even though our formulation of Eq.(\ref{eq:Lsummass}) 
allows us to choose any flavor eigenstate for $\sigma$, we only consider the case of $\sigma = e$ in this paper.
For this case the combinations of the form
$\frac{V_{\sigma i}^\ast V_{\sigma j}}{V_{\sigma i} V_{\sigma j}^\ast}$
are written as,
\bea
\frac{V_{e 1}^\ast V_{e 2}}{V_{e 1} V_{e 2}^\ast}=e^{i \alpha_{21}} ,\quad
\frac{V_{e 2}^\ast V_{e 3}}{V_{e 2} V_{e 3}^\ast}=e^{i (\alpha_{31}- \alpha_{21}-2\delta)} ,\quad
\frac{V_{e 3}^\ast V_{e 1}}{V_{e 3} V_{e 1}^\ast}=e^{i (-\alpha_{31}+2\delta)}.
\eea
Furthermore, we explicitly write Eq.(\ref{eq:Lsummass}) for the electron family number ($\alpha=e$) to emphasize simplifications to the equation when $\sigma = \alpha = e$.
\begin{multline}
\langle\bq,e|L_{e}(t)|\bq,e\rangle =|V_{e1}^{}|^4
  \left(1-\frac{2m_1^2 \sin^2(E_1(\bq)t)}{E_1^2(\bq)}\right) \\
+|V_{e2}^{}|^4\left(1-\frac{2m_2^2 \sin^2(E_2(\bq)t)}{E_2^2(\bq)}\right)+|V_{e3}^{}|^4 
  \left(1-\frac{2m_3^2 \sin^2(E_3(\bq)t)}{E_3^2(\bq)} \right) \\
+|V_{e1}^{}|^{2} |V_{e2}^{}|^{2} \left\{\left(1 +\frac{|\bq|^2-m_1 m_2 \cos(\alpha_{21})}
  {E_1(\bq)E_2(\bq)} \right) \cos(E_1(\bq)-E_2(\bq))t\right. \\
+\left. \left(1 - \frac{|\bq |^2-m_1 m_2 \cos(\alpha_{21}) }
  {E_1(\bq)E_2(\bq)} \right) \cos (E_1(\bq)+E_2(\bq)) t \right \} \\
+|V_{e2}^{}|^{2} |V_{e3}^{}|^{2}\left\{ \left(1+ 
  \frac{|\bq|^2-m_2 m_3 \cos(-\alpha_{21}+\alpha_{31}-2\delta)}{E_2(\bq)E_3(\bq)}
  \right) \cos(E_2(\bq)-E_3(\bq)) t \right. \\
+\left. \left(1-\frac{|\bq |^2-m_2 m_3 \cos(-\alpha_{21}+\alpha_{31}-2\delta)}
  {E_2(\bq)E_3(\bq)} \right) \cos(E_2(\bq)+E_3(\bq))t \right \} \\
+|V_{e3}^{}|^{2}|V_{e1}^{}|^{2} \left\{ \left( 1 +
  \frac{|\bq|^2-m_3 m_1 \cos(2 \delta-\alpha_{31})}{E_3(\bq)E_1(\bq)} \right)
  \cos (E_3(\bq)-E_1(\bq))t  \right. \\
+\left. \left(1 - \frac{|\bq |^2-m_3 m_1 \cos(2 \delta-\alpha_{31}) }{E_3(\bq)E_1(\bq)}
  \right)\cos(E_3(\bq)+E_1(\bq))t \right \},
 \label{eq:Lee}
\end{multline}
where the relevant elements of the PMNS matrix in Eq.(\ref{eq:PMNSmatrix}) are,
\bea
&&V_{e1}=c_{12}c_{13},\quad
V_{e2}=s_{12}c_{13}e^{i \frac{\alpha_{21}}{2}}, \quad
V_{e3}=s_{13} e^{-i \delta}e^{i\frac{\alpha_{31}}{2}}.
\label{eq:PMNSVe}
\eea
Those elements are important in the study of the short time ($m_i t \ll 1$) behavior of Eq.(\ref{eq:Lee}) as follows,
\bea
\langle\bq,e|L_{e}(t)|\bq,e\rangle &\simeq& 1-|m_{ee}|^2 t^2-(m^\dagger m)_{ee} t^2,
\label{eq:approxLee1} \\
&=&1-(2 |m_{ee}|^2+|m_{e\mu}|^2+|m_{e\tau}|^2) t^2,
\label{eq:approxLee2}
\eea
where we have used the common notation of,
\bea
|m_{ee}|&\equiv&|\sum_{i=1}^{3} V_{e i}^2 m_i |, \nn \\
&=&|m_1 c_{13}^2 c_{12}^2 +m_2 c_{13}^2 s_{12}^2 e^{i \alpha_{21} }+m_3 s_{13}^2 e^{i (\alpha_{31}-2 \delta)}|, \label{eq:mee} \\
(m^\dagger m)_{ee}&\equiv&\sum_{i=1}^3 |V_{ei}|^2 m^2_i, \nn \\
&=& {m_1}^2 c_{13}^4 c_{12}^4 +{m_2}^2 c_{13}^4 s_{12}^4+{m_3}^2 s_{13}^4.
\label{eq:mme}
\eea
In Eq.(\ref{eq:approxLee1}), the term $|m_{ee}|$ depends on Majorana phases from the elements of Eq.(\ref{eq:PMNSVe}) as shown in Eq.(\ref{eq:mee}).  While the term
$(m^\dagger m)_{ee}$ is independent of them as in Eq.(\ref{eq:mme}).
Thus, the short time behavior in Eq.(\ref{eq:approxLee2}) can be understood as follows.
The transition probability for $\nu_e \to \overline{\nu_\alpha}$ during the time interval $t$ is given as $|m_{\alpha e} t|^2$. During such time 
the electron number decreases as $\Delta L_e = -2 |m_{ee} t |^2-\sum_{\alpha=\mu}^{\tau} |m_{\alpha e} t|^2$.

Similarly the muon number ($\alpha=\mu$) can be written from Eq.(\ref{eq:Lsummass}) as,
\begin{multline}
\langle\bq,e|L_{\mu}(t)|\bq,e\rangle = |V_{\mu1}^{\ast}|^2 |V_{e1}^{}|^2
  \left(1-\frac{2m_1^2 \sin^2(E_1(\bq)t)}{E_1^2(\bq)}\right)\\
+|V_{\mu 2}^{\ast}|^2 |V_{e2}^{}|^2 
  \left(1-\frac{2m_2^2 \sin^2(E_2(\bq)t)}{E_2^2(\bq)}\right) 
+|V_{\mu 3}^{\ast}|^2 |V_{e3}^{}|^2
  \left(1-\frac{2m_3^2 \sin^2(E_3(\bq)t)}{E_3^2(\bq)}\right)\\
+\tmop{Re}(V_{\mu 1}^{\ast}V_{e 1}^{}V_{\mu 2}^{}V_{e 2}^{\ast})
  \left\{ \left( 1 + \frac{|\bq|^2-m_1 m_2 \cos(\alpha_{21})}{E_1(\bq)E_2(\bq)}\right)
  \cos(E_1(\bq)-E_2(\bq))t  \right.\\
+\left. \left(1 - \frac{|\bq|^2-m_1 m_2 \cos(\alpha_{21})}{E_1(\bq)E_2(\bq)}\right)
  \cos(E_1(\bq)+E_2(\bq))t \right \}\\
+\tmop{Re}(V_{\mu 2}^{\ast}V_{e 2}^{}V_{\mu 3}^{}V_{e 3}^{\ast})
  \left\{ \left( 1 + \frac{| \bq |^2-m_2 m_3 \cos(-\alpha_{21}+\alpha_{31}-2\delta)}
  {E_2(\bq)E_3(\bq)}\right) \cos(E_2(\bq)-E_3(\bq))t \right. \\
+\left.\left(1 - \frac{|\bq|^2-m_2 m_3 \cos(-\alpha_{21}+\alpha_{31}-2\delta)}
  {E_2(\bq)E_3(\bq)}\right) \cos(E_2(\bq)+E_3(\bq))t \right \} \\
+\tmop{Re}(V_{\mu 3}^{\ast}V_{e 3}^{} V_{\mu 1}^{}V_{e 1}^{\ast})
  \left\{ \left(1 + \frac{|\bq|^2-m_3 m_1 \cos(2\delta-\alpha_{31})}{E_3(\bq)E_1(\bq)}
  \right) \cos(E_3(\bq)-E_1(\bq))t \right. \\
+\left.\left(1 - \frac{|\bq|^2-m_3 m_1 \cos(2\delta-\alpha_{31})}{E_3(\bq)E_1(\bq)}
  \right) \cos(E_3(\bq)+E_1(\bq))t \right \}, \\
-\tmop{Im}(V_{\mu 1}^{} V_{e 1}^{\ast} V^{\ast}_{\mu 2} V_{e2}^{})
  \Biggl\{ - \sin(\alpha_{21}) \frac{m_1 m_2 (\cos (E_{1}(\bq) - E_{2}(\bq)) t
  -\cos(E_{1}(\bq)+E_{2}(\bq))t)}{E_{1}(\bq) E_{2}(\bq)} \\ 
-\sin(-\alpha_{21}+\alpha_{31}-2\delta)
  \frac{m_2 m_3 (\cos(E_{2}(\bq) - E_{3}(\bq)) t - \cos(E_{2}(\bq) + E_{3}(\bq)) t)}
  {E_{2}(\bq) E_{3}(\bq)} \\ 
-\sin(2\delta-\alpha_{31}) \frac{m_3 m_1 (\cos(E_{3}(\bq)-E_{1}(\bq)) t 
  -\cos (E_{3}(\bq) + E_{1}(\bq)) t)}{E_{3}(\bq) E_{1}(\bq) } \\
+\left(\frac{|\bq|}{E_1 (\bq)} -\frac{|\bq|}{E_2(\bq)} \right) \sin (E_{1}(\bq) +
  E_{2}(\bq)) t + \left( \frac{|\bq |}{E_1 (\bq)} +\frac{|\bq |}{E_2(\bq)} \right)
  \sin(E_{1}(\bq)-E_{2}(\bq)) t ) \\
+\left(\frac{|\bq |}{E_2 (\bq)} -\frac{|\bq |}{E_3(\bq)} \right) \sin (E_{2}(\bq) +
  E_{3}(\bq)) t + \left( \frac{|\bq |}{E_2(\bq)} +\frac{|\bq |}{E_3(\bq)} \right)
  \sin(E_{2}(\bq) - E_{3}(\bq)) t ) \\
+\left(\frac{|\bq |}{E_3(\bq)} -\frac{|\bq |}{E_1(\bq)} \right) \sin (E_{3}(\bq) +
  E_{1}(\bq)) t + \left( \frac{|\bq |}{E_3(\bq)} +\frac{|\bq |}{E_1(\bq)} \right)
  \sin(E_{3}(\bq) - E_{1}(\bq))t) \Biggl\}. \\
 \label{eq:muonnumber}
\end{multline}
In addition to the PMNS elements of Eq.(\ref{eq:PMNSVe}), the following elements are also relevant to Eq.(\ref{eq:muonnumber}),
\bea
&& V_{\mu1}^{}=-s_{12} c_{23}-c_{12}s_{23}s_{13} e^{i \delta}, \
V_{\mu2}^{}=(c_{12}c_{23}-s_{12}s_{23}s_{13} e^{i \delta}) e^{i\frac{\alpha_{21}}{2}}, \
V_{\mu3}^{}=s_{23}c_{13} e^{i\frac{\alpha_{31}}{2}}.
\eea
Selected PMNS coefficients in Eq.(\ref{eq:muonnumber}) are,
\bea
&&\rm{Re}(V_{e2}^{}V^{\ast}_{e3} V^{\ast}_{\mu2}V_{\mu3}^{})=s_{12}c_{13}^2s_{13}s_{23}
(c_{12} c_{23} \cos \delta-s_{12} s_{23} s_{13}) ,\\
&&\rm{Re}(V_{\mu 1}^{}V^{\ast}_{\mu 3} V_{e 1}^{\ast} V^{}_{e 3})= c_{12}c_{13}^2s_{13}s_{23}
(-s_{12} c_{23} \cos \delta-c_{12} s_{23} s_{13}),\\
&& \rm{Re}(V_{\mu 1}^{}V^{\ast}_{\mu 2} V_{e2}^{}V^{\ast}_{e1})
=c_{12} s_{12} c_{13}^2(-s_{12}c_{12}c_{23}^2-\cos 2\theta_{12} s_{23}s_{13} c_{23} \cos \delta
+c_{12}s_{12}s_{23}^2s_{13}^2),\\
&&\rm{Im}(V_{\mu 1}^{}V^{\ast}_{\mu 2} V_{e2}^{}V^{\ast}_{e1})=-
c_{12} c_{13}^2 s_{13} s_{12} s_{23} c_{23} \sin \delta.  \label{eq.Ja}
\eea
Here we find the imaginary term of Eq.(\ref{eq.Ja}) is proportional to the Jarlskog invariant
\cite{Jarlskog:1985ht}.
The time dependent part with the  coefficient given by the Jarlskog invariant is also sensitive to CP violation of the Majorana phases.
By taking the small momentum limit  in Eq.(\ref{eq:muonnumber}), we find the same terms as the last term in Eq.(\ref{eq:zerolimit}) with $\alpha=\mu$ and $\sigma=e$. Excluding the Jarlskog invariant, 
they are  written as,
\bea
&& \sum_{\{ i,j\}}\left[\cos{((m_i-m_j)t)}-\cos{((m_i+m_j)t)}\right] 
      \tmop{Im}\left[\frac{V_{\sigma i}^{\ast}V_{\sigma j}^{}}
      {V_{\sigma i}^{}V_{\sigma j}^{\ast}} \right]= 2 \{  \sin m_1 t \sin m_2 t \sin(\alpha_{21}) \nn \\
&&+  \sin m_2 t \sin m_3t \sin (-\alpha_{21}+\alpha_{31}-2 \delta)+
\sin m_3t \sin m_1t \sin(2 \delta-\alpha_{31}) \}.
\eea 
Lastly, tauon lepton number $(\alpha=\tau)$ can be obtained by replacing $V_{\mu i}$  
with $V_{\tau i}$ in Eq.(\ref{eq:muonnumber}).

We prepare Eq.(\ref{eq:expectationresult}) for numerical calculations by considering a dimensionless time denoted as $\tau$
\bea
t[1/\mbox{\rm eV}]&=& \frac{\tau}{0.02 [\mbox{\rm eV}]}, \nn \\
t[\mbox{picosec}] &=& 3.3 \times 10^{-2} \tau,
\eea
where $0.02$[eV] is a reference momentum.  This allows for the momentum of the neutrino, $|\bq|$[eV], to be expressed by a dimensionless ratio with the reference momentum,
\bea
\hq=\frac{|\bq|}{0.02}.
\eea
Furthermore, we also introduce a dimensionless energy as,
\bea
\hEi=\sqrt{1+\frac{m^2_i}{|\bq|^2}}.
\eea
From the combination of the three quantities $\tau$, $\hq$, and $\hEi$ we write the time evolution factors from Eq.(\ref{eq:Lee}) and Eq.(\ref{eq:muonnumber}) as,
\bea
\cos(E_i(\bq)\pm E_j(\bq))t &=& \cos(\hEi\pm\hEj)\hat{q} \tau, \\
\sin(E_i(\bq)\pm E_j(\bq))t &=& \sin(\hEi\pm\hEj)\hat{q} \tau.
\eea   
This results in an overall equation similar to that of Eq.(\ref{eq:expectationresult}),
\begin{multline}
    \langle\bq,\sigma|L_{\alpha}(t)|\bq,\sigma\rangle=
      \sum_i \left|V_{\alpha i}^{}\right|^2\left|V_{\sigma i}^{}\right|^2
      \left(\frac{1+\cos{(2\hat{E_i} \hq \tau)}}{\hat{E_i}^2}+
      \cos{(2 \hat{E_i} \hq \tau)}\right) \\
    +\sum_{\{i,j\}}\tmop{Re}\left[V_{\alpha i}^{\ast}V_{\sigma i}^{}
      V_{\alpha j}^{}V_{\sigma j}^{\ast}\right] \Biggl[\cos{((\hat{E_i}-\hat{E_j})\hq \tau)}+
      \cos{((\hat{E_i}+\hat{E_j})\hq \tau)} \Biggr. \\
    +\Biggl.\left(\cos{((\hat{E_i}-\hat{E_j})\hq \tau)}-
      \cos{((\hat{E_i}+\hat{E_j})\hq\tau)}\right)\left(\frac{1-\sqrt{\hat{E_i}^2-1}
      \sqrt{\hat{E_j}^2-1}\tmop{Re}\left[\frac{V_{\sigma i}^{\ast}V_{\sigma j}^{}}
      {V_{\sigma i}^{}V^{\ast}_{\sigma j}}\right]}{\hat{E_i}\hat{E_j}} \right) \Biggr] \\
    -\tmop{Im}\left[V_{\alpha 1}^{\ast}V_{\sigma 1}^{}V_{\alpha 2}^{}V_{\sigma 2}^{\ast}\right]
      \sum_{\{i,j\}}\Biggl[\left(\frac{1}{\hat{E_i}}-
      \frac{1}{\hat{E_j}}\right)\sin{((\hat{E_i}+\hat{E_j})\hq \tau)}+\left(\frac{1}{\hat{E_i}}+\frac{1}{\hat{E_j}}\right)\sin{((\hat{E_i}-\hat{E_j})\hq \tau)}\Biggr.\\
    \Biggl.-\frac{\sqrt{\hat{E_i}^2-1}\sqrt{\hat{E_j}^2-1}}{\hat{E_i}\hat{E_j}}
      \tmop{Im}\left[\frac{V_{\sigma i}^{\ast}V_{\sigma j}}
      {V_{\sigma i}V_{\sigma j}^{\ast}}\right]\Bigl(\cos{((\hat{E_i}-\hat{E_j})\hq\tau)}-
      \cos{((\hat{E_i}+\hat{E_j})\hq \tau)}\Bigr)\Biggr]
\label{eq:numericalequation}
\end{multline}
\subsection{Results}
\begin{figure}[ht!]
\begin{tabular}{cc}
\includegraphics[width=0.45\linewidth]{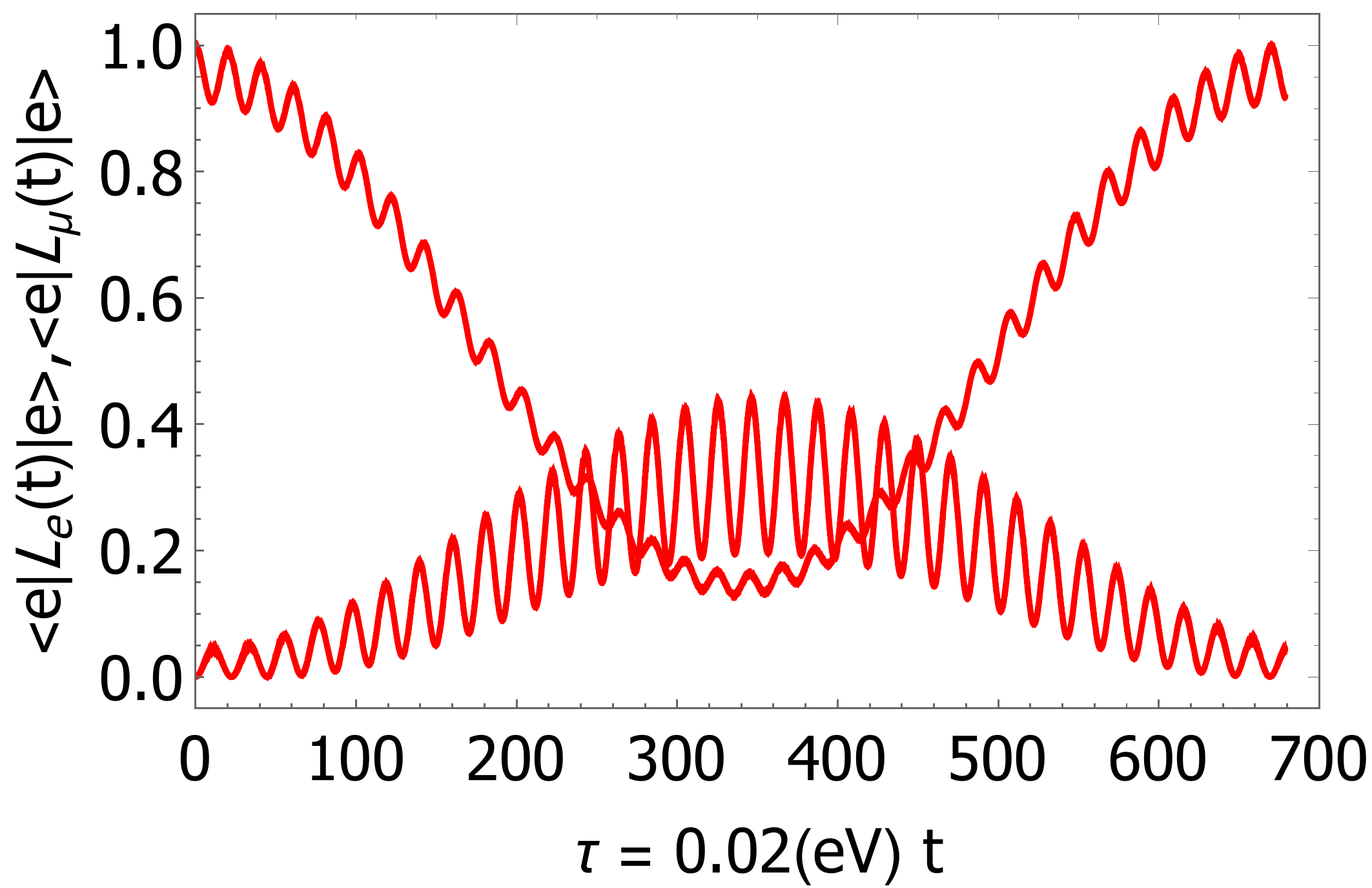}  
&
\includegraphics[width=0.45\linewidth]{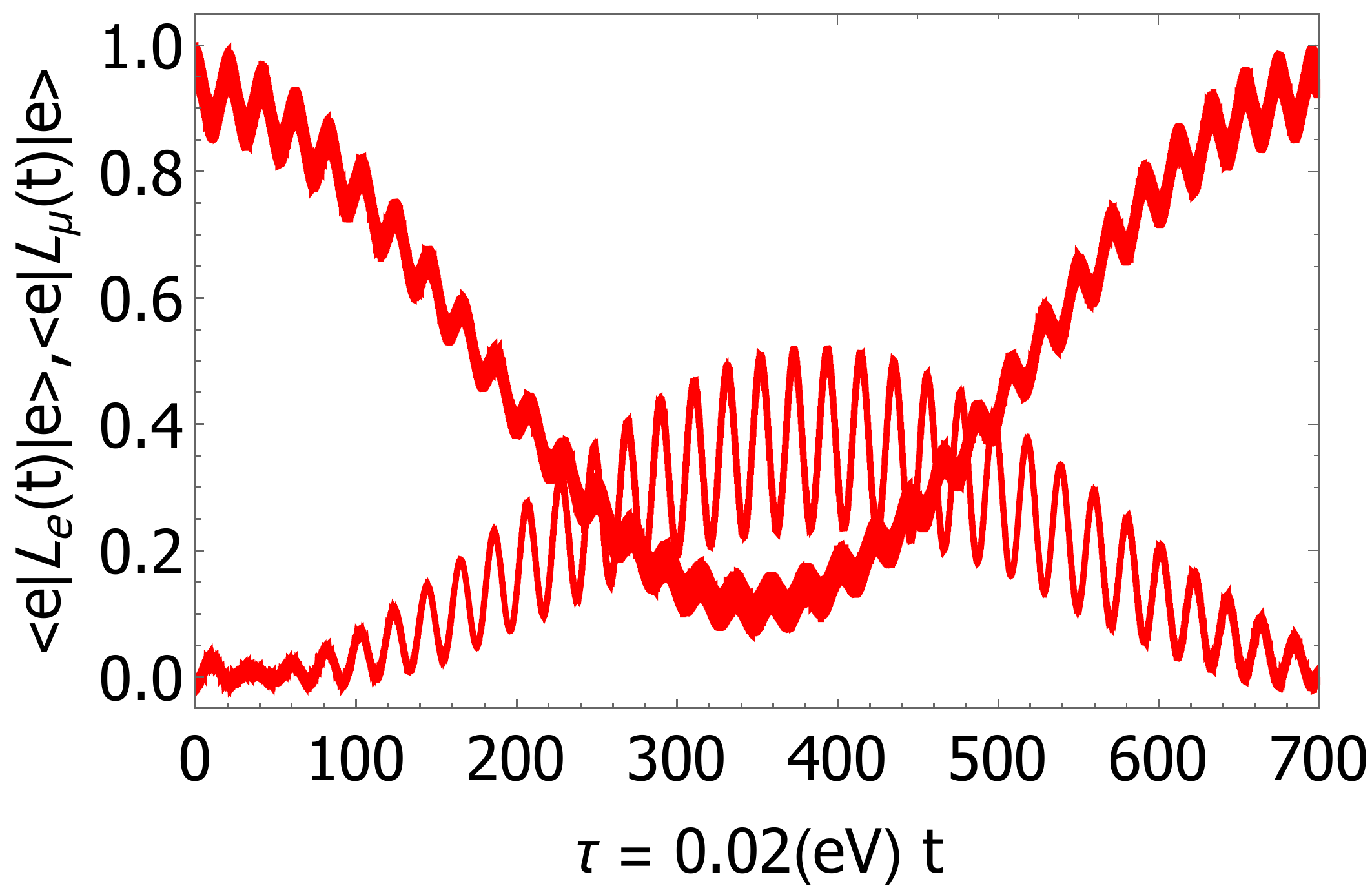}
 \\
Normal hierarchy case & Inverted hierarchy case 
\end{tabular}
	\caption{
Time dependence of lepton family numbers $\langle e|L_{e}(t)|e \rangle$ 
and $\langle e|L_\mu(t)| e \rangle$. 
 The lightest neutrino mass is $0.01$ (eV) and the momentum of the neutrinos are $|\bq|=0.2$ (eV).
 For the horizontal axis, 
we choose to use the dimensionless time $\tau=0.02 (\mbox{eV}) t$.   The curves which start at $(\tau=0,1.0)$ are the electron number $\langle e|L_e(t)| e \rangle$  and the ones  which start at  $(\tau=0, 0.0)$
are the muon number $\langle e|L_\mu(t)| e \rangle$.  Although the Majorana phases
are taken to be $(\alpha_{21}, \alpha_{31})= (\pi,\pi+2\delta)$ in the both figures, the other choice of the Majorana phases such as $(\alpha_{21}, \alpha_{31})=(0,2\delta)$ does not lead
to visible difference.
}
\label{Fig2}
\end{figure}

\begin{figure}[ht!]
\begin{tabular}{cc}
\includegraphics[width=0.45\linewidth]{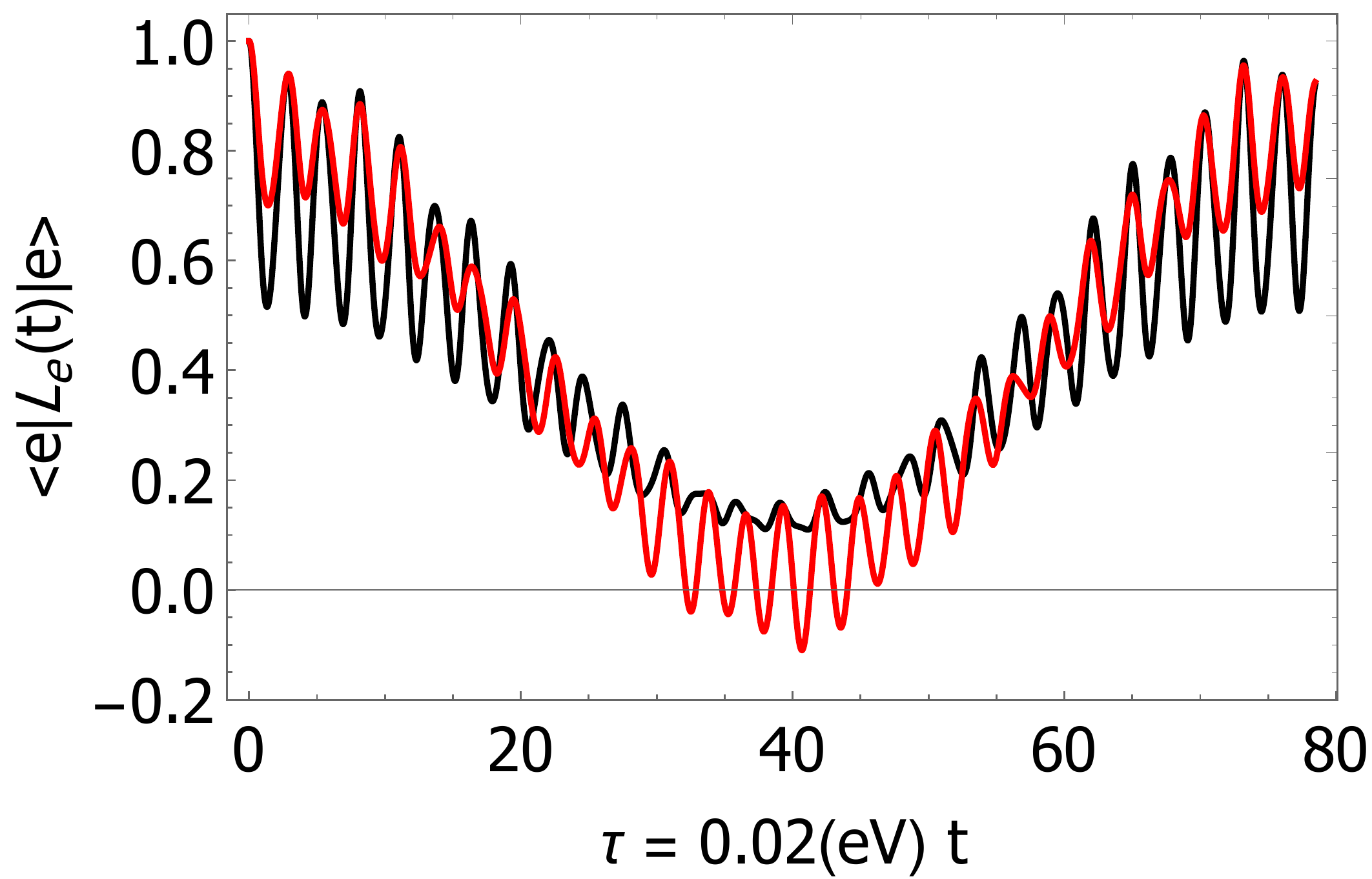}  
&
\includegraphics[width=0.45\linewidth]{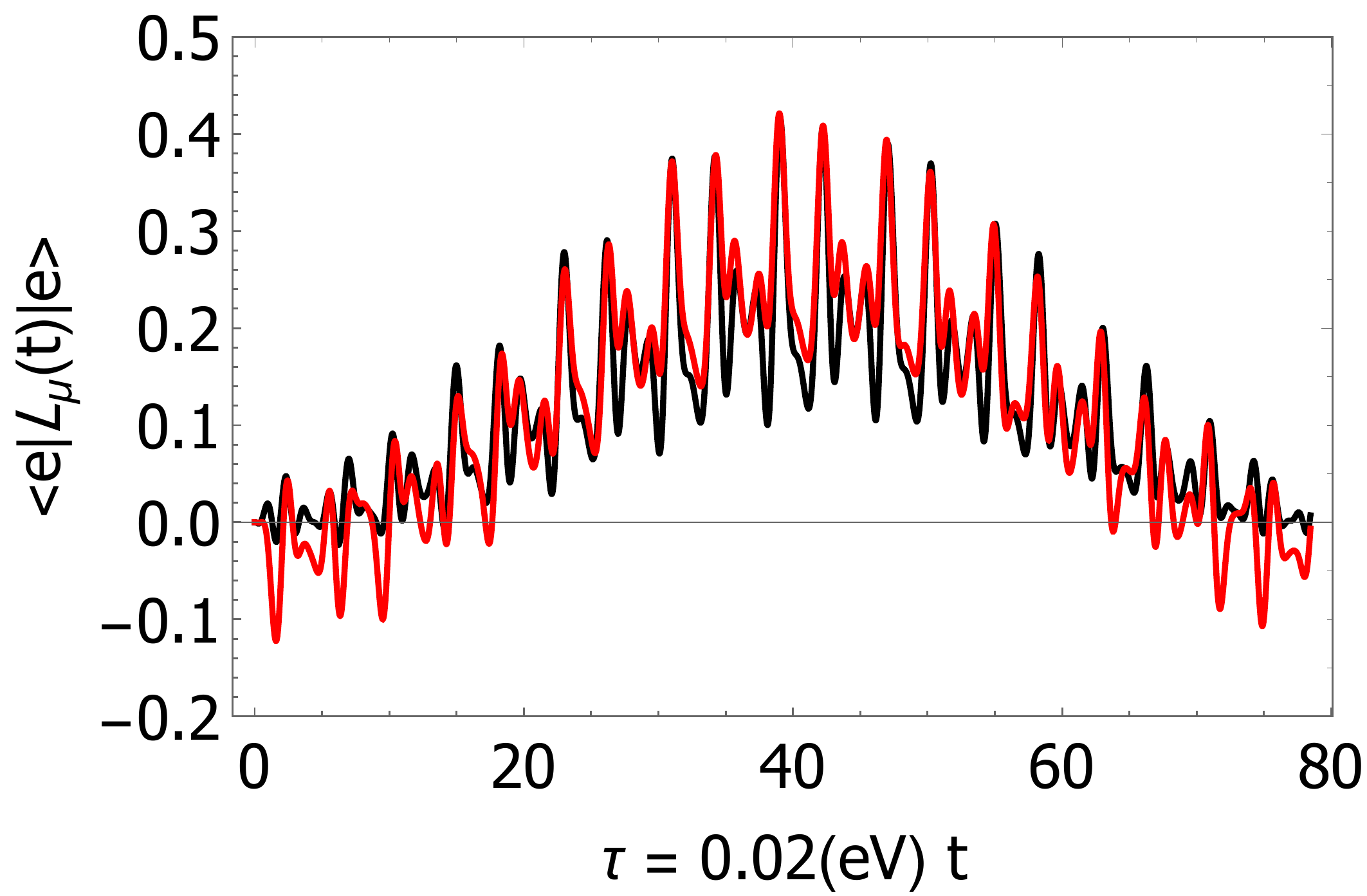}
\\
Electron number for normal hierarchy & Muon number for normal hierarchy 
\end{tabular}
\caption{
Time dependence of lepton family numbers $\langle e|L_{e}(t)|e \rangle$ (left figure)
and $\langle e|L_\mu(t)| e \rangle$ (right figure) for the  normal hierarchy case. 
The momentum of the neutrino is  $|\bq|=0.02$ (eV) and the other parameters are the same as those used for Fig.\ref{Fig2}.
The  black and red lines show the cases of Majorana phases
$(\alpha_{21},\alpha_{31})=(0,2\delta)$ and $(\pi,\pi+2\delta)$ 
respectively.
}
\label{Fig3}
\end{figure}

\begin{figure}[ht!]
\begin{tabular}{cc}
\includegraphics[width=0.45\linewidth]{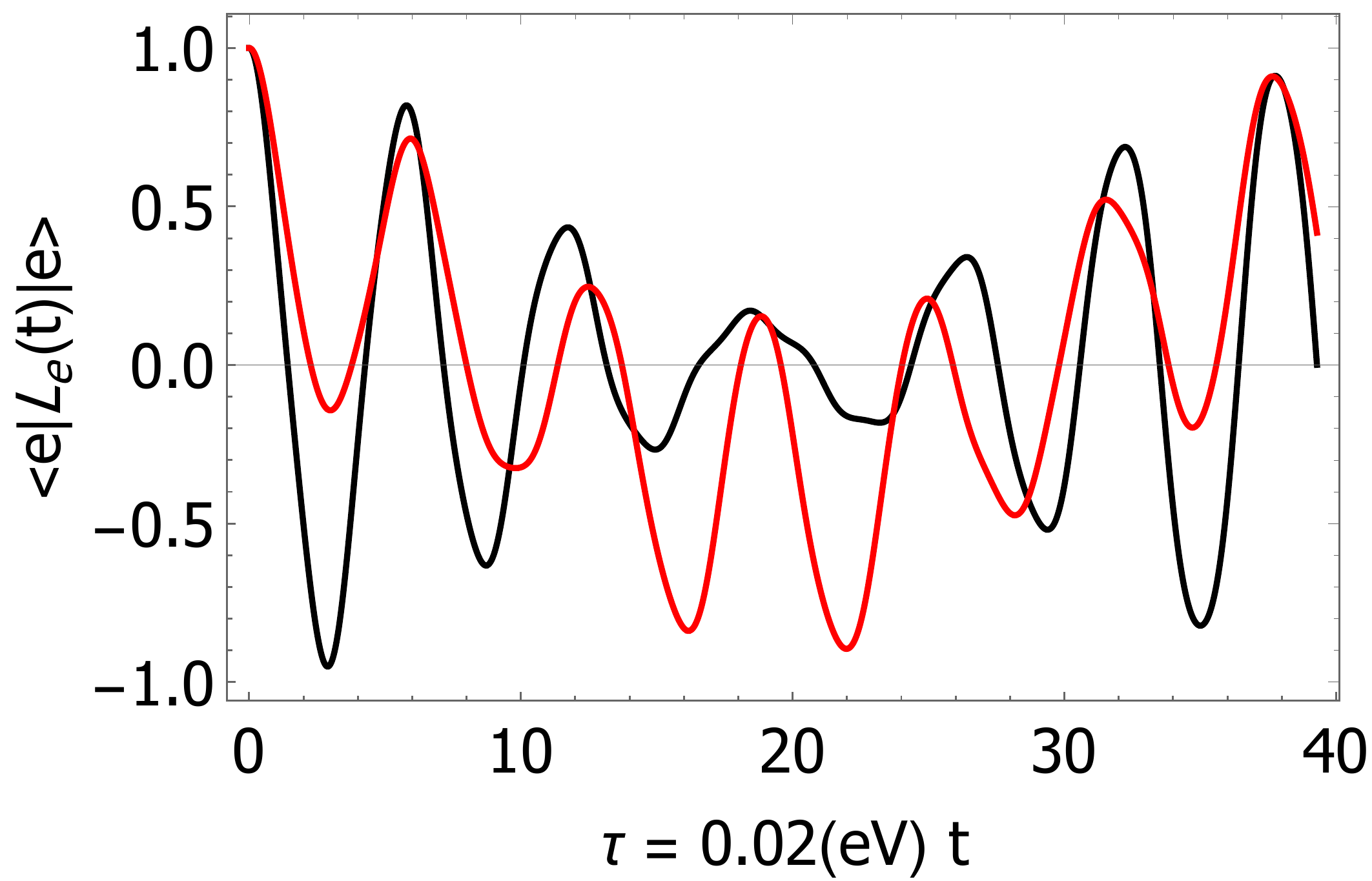}
& 
\includegraphics[width=0.45\linewidth]{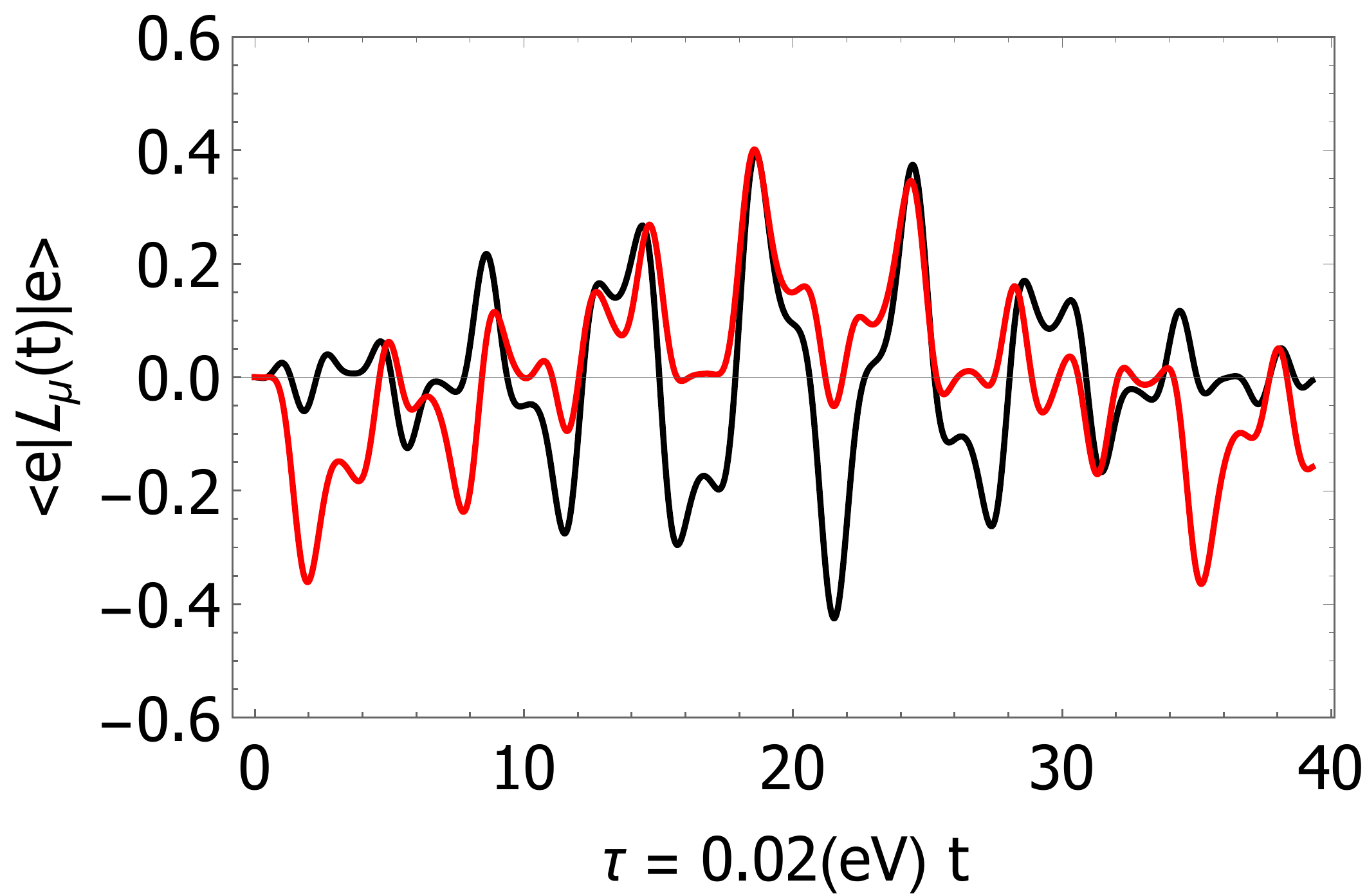}
\\
Electron number for normal hierarchy  & Muon number for normal hierarchy 
\end{tabular}
\caption{
Time dependence of  electron  family number $\langle e|L_{e}(t)|e \rangle$ (left figure)
and muon family  number $\langle e|L_\mu(t)| e \rangle$ (right figure) for the  normal hierarchy case.
The momentum of the neutrino is $|\bq|=0.0002$ (eV) and the lightest neutrino mass is $0.01$ (eV).
The black and red lines show the cases of Majorana phases
$(\alpha_{21},\alpha_{31})=(0,2\delta)$ and $(\pi,\pi+2\delta)$ 
respectively.
}
\label{Fig4}
\end{figure}

\begin{figure}[ht!]
\begin{tabular}{cc}
\includegraphics[width=0.475\linewidth]{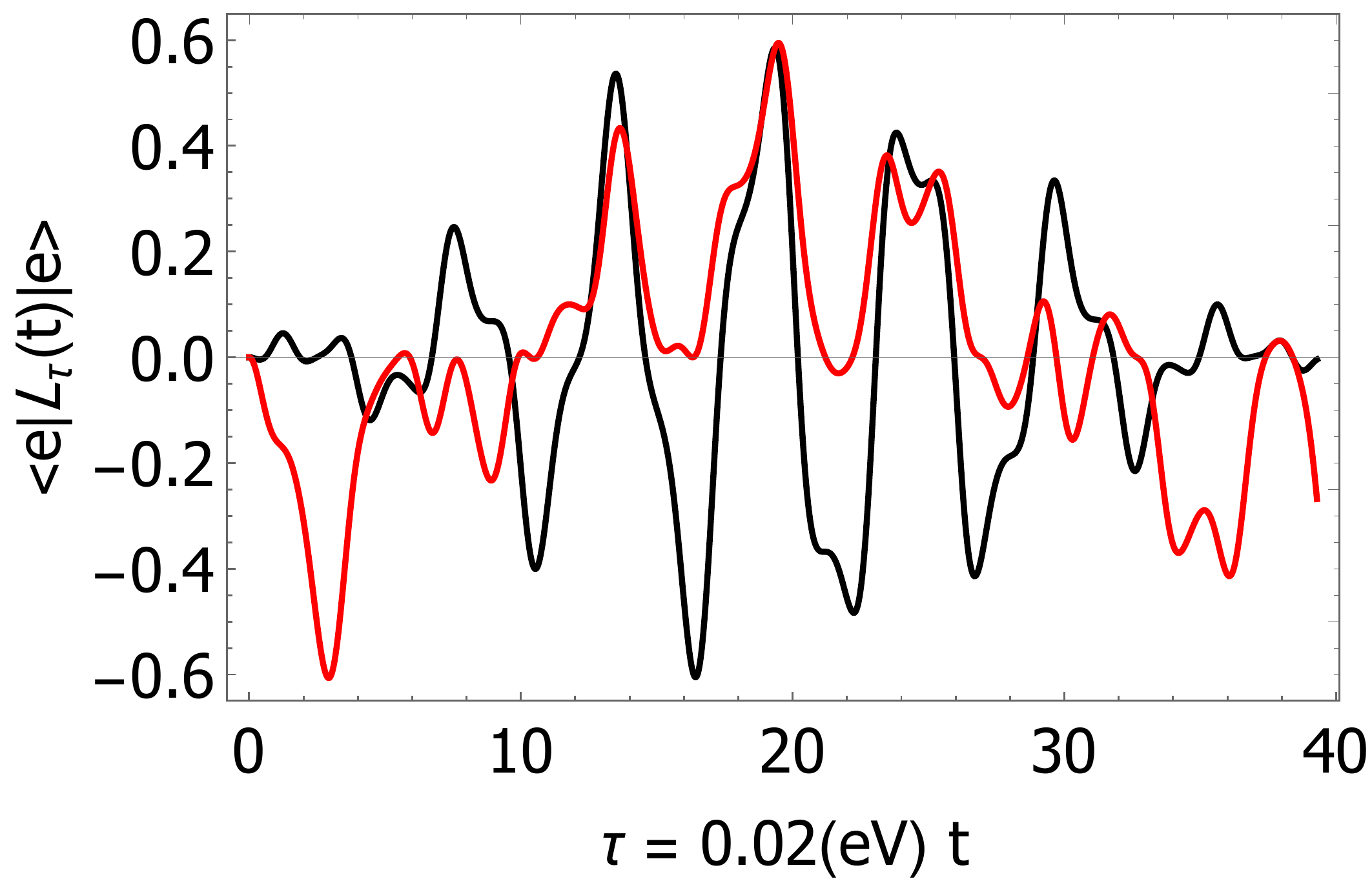}
&
\includegraphics[width=0.475\linewidth]{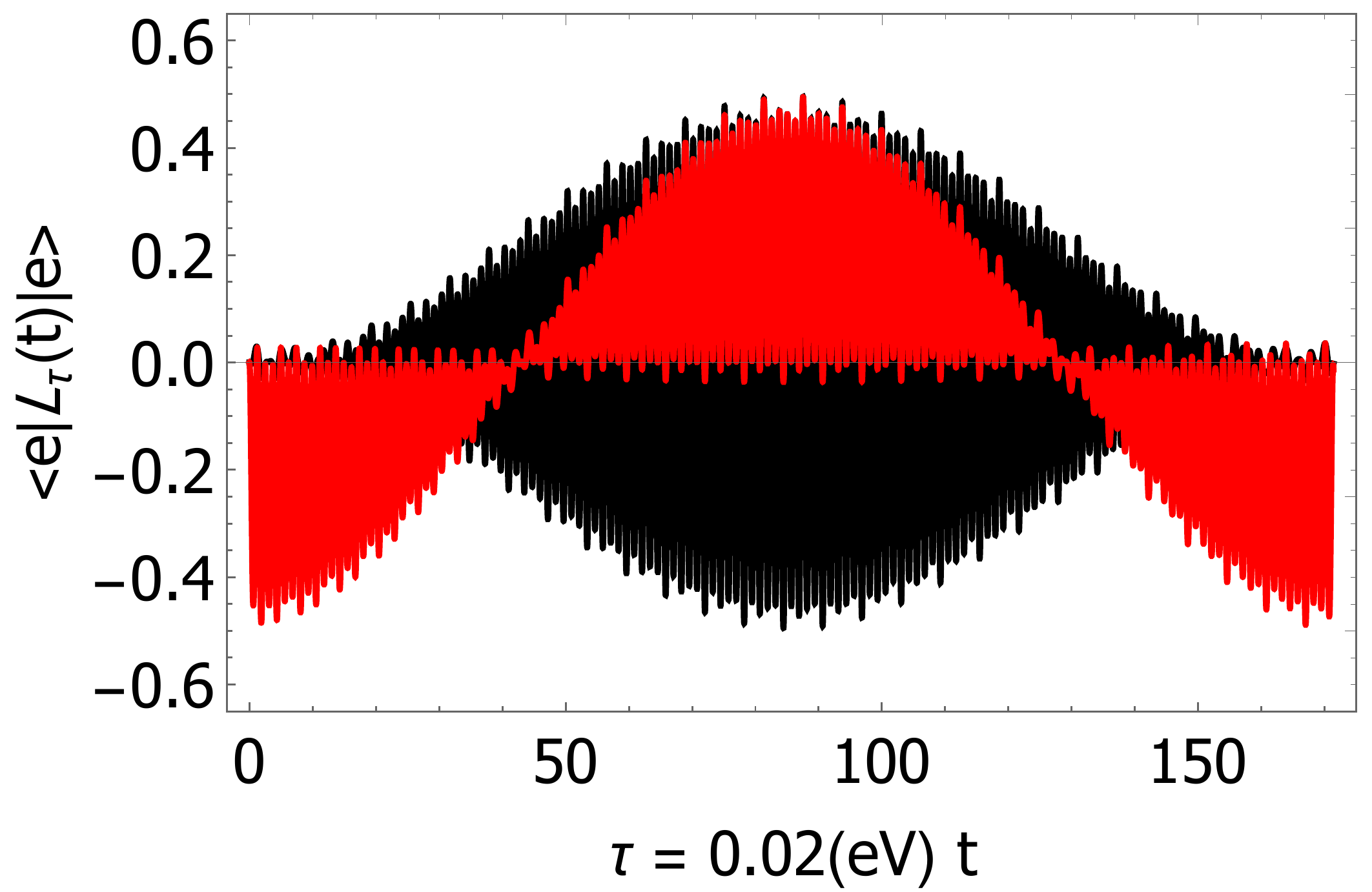} \\
Tauon number for normal hierarchy & Tauon number for inverted hierarchy
\end{tabular}
\caption{
Time dependence of $\tau$ lepton family number $\langle e|L_\tau(t)|e \rangle$
for normal hierarchy case (left) and for the inverted hierarchy case (right).
The lightest neutrino mass
is $0.01$ (eV) and the momentum of neutrino is $|\bq|=0.0002$ (eV).
The black  and red lines show the case
of Majorana phases $(\alpha_{21},\alpha_{31})=(0,2\delta)$ and $(\pi,\pi+2\delta)$.
\label{Fig5}
 }
\end{figure}

\begin{figure}[ht!]
\begin{center}
\includegraphics[width=0.475\linewidth]{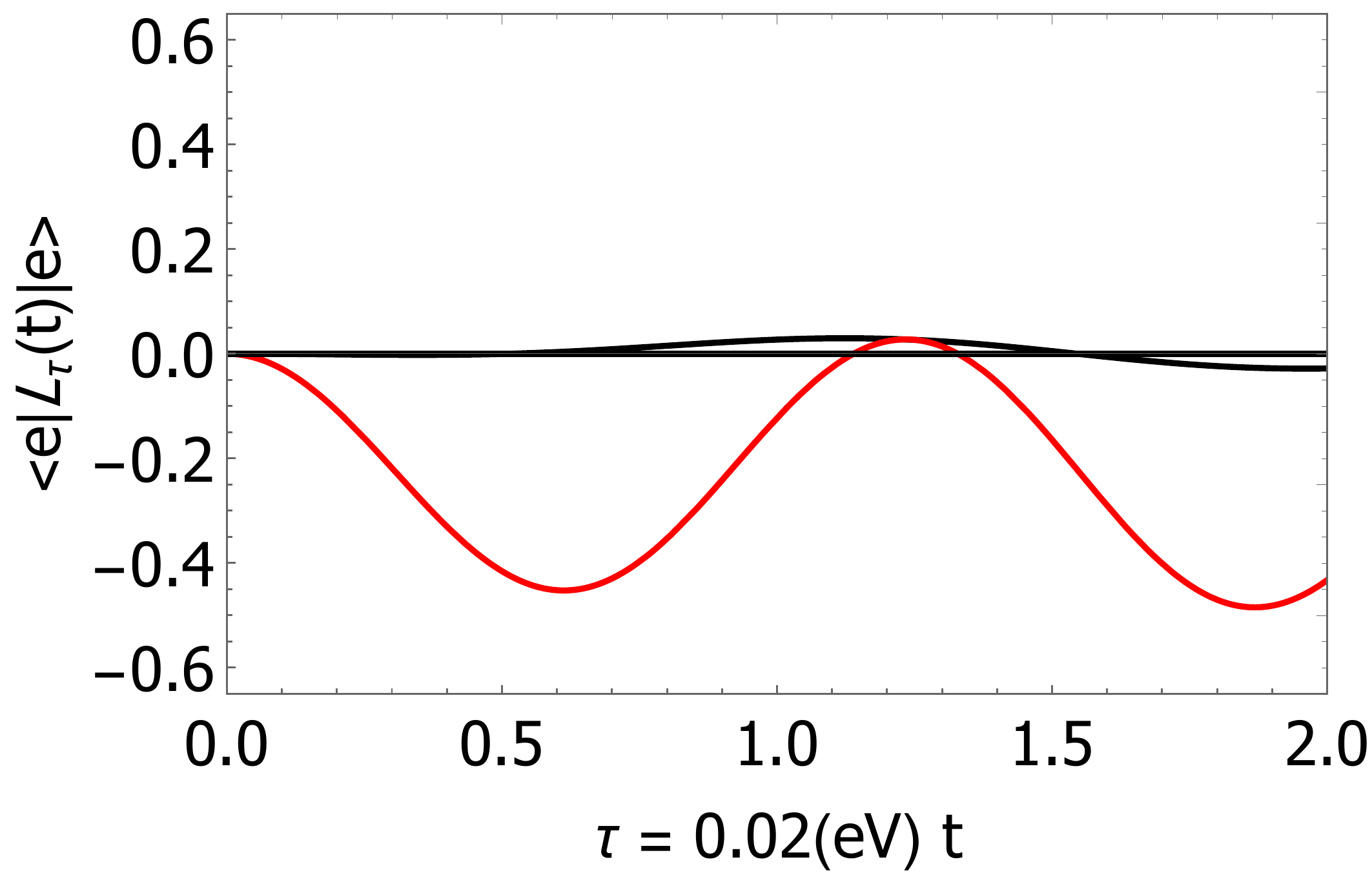} \\
\caption{
    We show the short time behavior of tauon  number $\langle e|L_\tau(t)|e \rangle$ for the inverted hierarchy case of the right panel of Fig.\ref{Fig5}.
    The black  and red lines show the case of Majorana phases $(\alpha_{21},\alpha_{31})=(0,2\delta)$ and $(\pi,\pi+2\delta)$ respectively. 
\label{Fig6}
}
\end{center}
\end{figure}

In Fig.\ref{Fig2}-\ref{Fig6}, the expectation values are plotted as functions
of $\tau$. 
In the numerical calculations we adopt the following data from Ref.\cite{Esteban:2020cvm}.
For the normal mass hierarchy case we use $\Delta m^2_{21}=7.42 \times 10^{-5}$(eV$^2$) and $\Delta m^2_{31}=2.517 \times 10^{-3}$(eV$^2$). The Dirac phase and sine of the mixing angles
are chosen as
$\delta=1.09444\pi$, $s_{12}=0.551362$, $s_{13}=0.148963$, and $s_{23}=0.756968$.
For the inverted mass hierarchy case we adopt 
$\Delta m^2_{21}=7.42 \times 10^{-5}$(eV$^2$) and $\Delta m^2_{23}=2.498 \times 10^{-3}$(eV$^2$). The Dirac phase and sine of the mixing angles
are chosen as
$\delta=1.56667\pi$, $s_{12}=0.551362$, $s_{13}=0.149599$, and $s_{23}=0.758288$.
We assume the lightest neutrino mass to be $0.01$[eV].  From that choice of the lightest neutrino mass each mass eigenvalue $m_i$[eV] is given by,
\bea
m_1&=&0.0100, \quad m_2=0.0132,  \quad  m_3=0.0512, \quad (\mbox{Normal}), 
\label{eq:massnormal} \\
m_1&=&0.0502,  \quad m_2=0.0510,  \quad  m_3=0.0100, \quad (\mbox{Inverted}). 
\label{eq:massinverted}
\eea
We choose the following values for Majorana phases
$(\alpha_{21},\alpha_{31})=(0,2\delta)$ and $(\pi,\pi+2\delta)$.
The former corresponds to the maximum value for $|m_{ee}|$ in Eq.(\ref{eq:mee}) and the latter corresponds to the minimum value for $|m_{ee}|$ respectively.  We explicitly write the minimum and maximum values for the different mass hierarchies,
\bea
0.00175 \le |m_{ee}| \le 0.0119&& \quad (\mbox{Normal}), \\
0.0182 \le |m_{ee}| \le 0.0496 && \quad (\mbox{Inverted}).
\eea
\newpage
In Fig.\ref{Fig2}, we show the time evolution of  electron number and muon number for an electron neutrino with the momentum $|\bq|=0.2$[eV] as an initial state. There the momentum is larger than neutrino masses in Eq.(\ref{eq:massnormal}) and Eq.(\ref{eq:massinverted}).  
For this case the coefficients of  $\cos(E_i+E_j)t$ and $\sin(E_i+E_j)t$, from Eq.(\ref{eq:Lee}) and Eq.(\ref{eq:muonnumber}), are suppressed compared to those of $\cos(E_i-E_j) t$ and $\sin(E_i-E_j)t$.
This is because of the relativistic nature of the neutrino for this case.
%

In addition, Fig.\ref{Fig2} shows the lepton family number oscillating with a long period and a
large amplitude. The long period, which we denote as $\tau_L$, originates from the smallest energy difference of $E_2-E_1$.
\bea
\tau_L \equiv \frac{2\pi}{E_2-E_1} \times 0.02(\mbox{eV})= 679 (\mbox{Normal}), \quad 699 (\mbox{Inverted}).
\label{eq:tauL}
\eea
There are also short periods, rapid oscillations, with small amplitudes.
In the normal hierarchical case the short periods, denoted as $\tau_{S1}$ and $\tau_{S2}$, are given by,
\bea
\tau_{S1} \equiv \frac{2\pi}{E_3-E_1} \times 0.02({\rm eV})=20.3,  \quad
\tau_{S2} \equiv \frac{2\pi}{E_3-E_2} \times 0.02 ({\rm eV})=20.9.
\eea
In the inverted hierarchical case the short periods are given by,
\bea
\tau_{S1} \equiv \frac{2\pi}{E_2-E_3} \times 0.02({\rm eV})=20.5,  \quad
\tau_{S2} \equiv \frac{2\pi}{E_1-E_3} \times 0.02({\rm eV})=21.1.
\eea
The rapid oscillation shows the behavior of the beat.  The period of the beat in the dimensionless time is given as,
\bea
\tau_{\rm beat} \equiv \frac{2 \pi}{E_2-E_1} \times 0.02 ({\rm eV})=\tau_L .
\label{taubeat}
\eea
where $\tau_{\rm beat}$ is the same as $\tau_L$ in Eq.(\ref{eq:tauL}) as can be seen
in Fig.\ref{Fig2}. 
Next we discuss the difference in the change of the amplitude between beat of the muon family number and that of the electron family number.
From Eq.(\ref{eq:Lee}) of the electron family number the terms proportional to $\cos(E_3-E_2) t$ and $\cos(E_3-E_1) t$ are suppressed by a factor of $s_{13}^2$.
While for the muon family number in Eq.(\ref{eq:muonnumber}) the coefficients 
of $\sin(E_3-E_2) t$ and $\sin(E_3-E_1) t$ are proportional to $s_{13}$.
This results in a larger amplitude for the beat of the muon number than that of the electron number.  Lastly, we note that both lepton family numbers lie in the range $[0,1]$
as one can expect from  Eq.(\ref{eq:ultralimit}). 

In Fig.\ref{Fig3}, the momentum is taken to be the value $|\bq|=0.02$[eV], which is comparable with the lightest neutrino mass of Eq.(\ref{eq:massnormal}) and Eq.(\ref{eq:massinverted}).
In contrast to Fig.\ref{Fig2}, where the lepton family numbers are always positive, the muon family number of Fig.\ref{Fig3} can take both negative and positive values; the electron family number can also take negative values for the specific choice of Majorana phases $(\alpha_{21},\alpha_{31})=(\pi, \pi+2 \delta)$. However, most of the time they are positive. 
We also note that the dependence on the Majorana phases 
are visible in this case.  In  the left of Fig.\ref{Fig3}, the electron family number sharply
decreases for the case $(\alpha_{21},\alpha_{31})=(0, 2 \delta)$ where $|m_{ee}|$ in 
Eq.(\ref{eq:mee}) is large. This is consistent with the short time behavior of
$ \langle e|L_e(t) |e \rangle $ as shown in Eq.(\ref{eq:approxLee1}).

In Fig.\ref{Fig4} we further lower the momentum to $|\bq|=0.0002$[eV], which is comparable with
the relic neutrino temperature for massless neutrinos $T_{relic}=(\frac{4}{11})^\frac{1}{3} \simeq1.9$[K].
Since even the lightest mass of Eq.(\ref{eq:massnormal}), $m_1=0.01$[eV], is larger than the momentum, all the neutrinos are non-relativistic.
In this figure, we can see the lepton family numbers are more sensitive to Majorana phases compared to the case in Fig.\ref{Fig3} where the momentum is larger. 
In addition, both the muon family number and electron family number spend equal time with positive and negative values.
The dominant contribution to the electron family number can be approximated as,
\begin{multline}
\lim_{|\bq|\rightarrow 0_+} \langle\bq,e|L_e(t)|\bq,e\rangle \bigl|_{s_{13}=0}\bigr. \simeq \\
4s_{12}^2 c_{12}^2\frac{1}{2} \left \{ 
\sin^2(\frac{\alpha_{21}}{2}) \cos(m_- t) + \cos^2(\frac{\alpha_{21}}{2})  
\cos(m_+ t) + \cos(m_+ t) \cos(m_- t) \right \} \\
+(c_{12}^2-s_{12}^2)  \sin(m_+ t ) \sin(m_- t)+
(c_{12}^2-s_{12}^2)^2 \cos(m_+ t) \cos(m_- t) ,
  \label{eq:approxLee}
\end{multline}
where $m_\pm=m_2 \pm m_1$.
If we take the maximal mixing limit of $c^2_{12}=s^2_{12}=\frac{1}{2}$ Eq.(\ref{eq:approxLee}) becomes,
\begin{equation}
  \begin{aligned}
  \lim_{|\bq|\rightarrow 0_+} \langle\bq,e|L_e(t)|\bq,e\rangle \simeq&
  \frac{1}{2} \left(\cos(m_+ t) \cos(m_- t)+ \sin^2(\frac{\alpha_{21}}{2})
  \cos (m_- t) + \cos^2(\frac{\alpha_{21}}{2} )  \cos (m_+ t) \right) \\
  =&\begin{cases}
    \cos(m_+  t)  \cos^2(\frac{m_-}{2}t) &
      \qquad \text{when } \alpha_{21}=0, \\
    \cos(m_-  t) \cos^2(\frac{m_+}{2}t) &
      \qquad \text{when } \alpha_{21}=\pi .
    \end{cases}
  \end{aligned}
  \label{eq:approxforsmallq}
\end{equation}
Eq.(\ref{eq:approxforsmallq}) qualitatively explains the peak-to-peak amplitude modulation
of the electron family numbers in the left figure of Fig.\ref{Fig4}. The black line 
has a slow peak-to-peak amplitude modulation and short period oscillation. Therefore, it changes 
the sign very frequently and at some time the amplitude vanishes.
Qualitatively this is case when $\alpha_{21}=0$ of Eq.(\ref{eq:approxforsmallq}).
The red line has a rapid modulation of the peak-to-peak amplitude and a long period oscillation.
This is case when $\alpha_{21}=\pi$ of Eq.(\ref{eq:approxforsmallq}).

In contrast to the electron family number, the muon family number in the right figure of Fig.\ref{Fig4} has a fine structure. In the vanishing limit of $s_{13}$ and low
momentum limit $|\bq| \rightarrow 0_+$, the muon family number can be approximately given as,
\begin{equation}
  \begin{split}
  \lim_{|\bq|\rightarrow 0_+} \langle \bq,e|L_\mu(t)|\bq,e \rangle
  &\simeq-4 (s_{12} c_{12} c_{23})^2 
    \biggl[\left(\sin\frac{\alpha_{21}}{2} \sin\frac{m_+ t}{2}\right)^2
    \cos (m_- t)\\
    &\phantom{XXXXXXXXXXXXXXXXX}
    +\left(\cos\frac{\alpha_{21}}{2} \sin\frac{m_- t}{2}\right)^2
    \cos(m_+ t)\biggr]\\
  &=-4 (s_{12} c_{12} c_{23})^2 \begin{cases}
    \cos(m_+ t)  \sin^2(\frac{m_-}{2}t)
      \qquad \text{when } \alpha_{21}=0, \\
    \cos(m_- t) \sin^2(\frac{m_+}{2}t)
      \qquad \text{when } \alpha_{21}=\pi.
    \end{cases}
  \end{split}
 \label{eq:approxforsmallqmuon}
\end{equation}
Where the approximate formula includes only two periods. The shorter period corresponds to
$\frac{2\pi}{m_+} \times 0.02 \sim 5.44 $ and the longer period corresponds to $\frac{2 \pi}{m_-} \times 0.02 \sim 39$.  The fine structure of the muon family number comes from
the effect of non-zero $V_{e3}$. The effect of the third mass eigenstate  adds the oscillation 
with the frequencies of $m_3 -m_2 < m_3-m_1 < m_3+m_1  <m_3 +m_2 $. They corresponds to periods from $1.95$ to $3.31$ in the dimensionless time unit.

\begin{figure}[ht!]
\begin{center}
\includegraphics[width=0.475\linewidth]{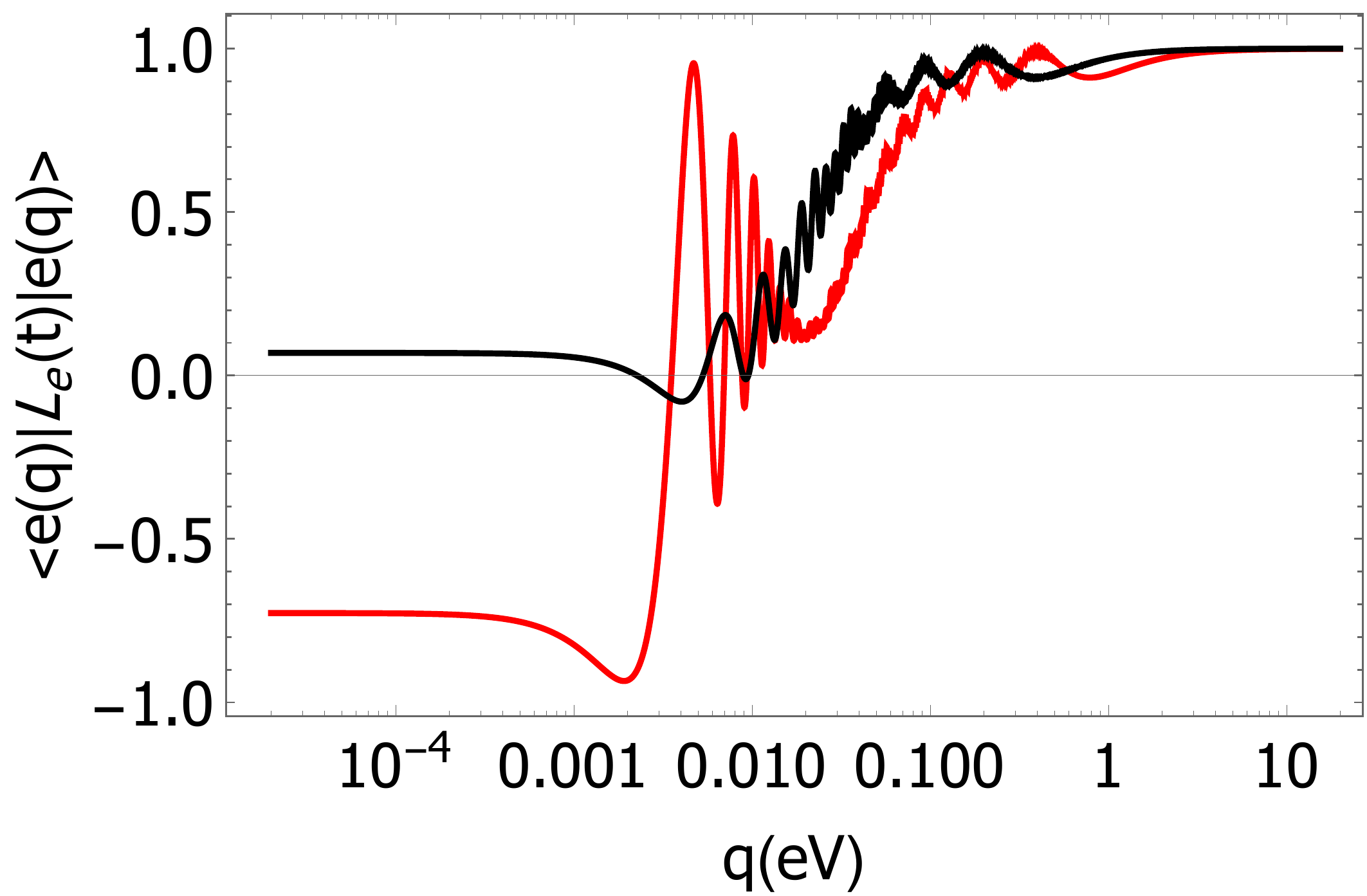} \\
\caption{
 The momentum dependence of  $\langle e(q)|L_e(t)|e(q) \rangle$ for two 
different fixed time $t=50(1/{\rm eV}) \tau$. 
The momentum range is from $2.0 \times 10^{-5}$(eV) to $20$ (eV).
The red curve shows the case with $\tau=40$ and the black curve corresponds to
$\tau=20$.   The Majorana phases are chosen as  $(\alpha_{21},\alpha_{31})=(0,2\delta)$. The figure corresponds to the normal hierarchy case.
\label{Fig7}
}
\end{center}
\end{figure}

In Fig.\ref{Fig5}, the momentum is the same as that of Fig.\ref{Fig4}. 
However, we study the tauon family number for the two different neutrino mass hierarchies, normal and inverted.  From the figure a striking difference between two hierarchies can been seen. We note even though the scales of the horizontal axis are different the effects of the hierarchies is present for equivalent scales.
In order to understand the difference caused by the hierarchies,
we first write the tauon lepton family number with 
the same approximation as that of muon family number in Eq.(\ref{eq:approxforsmallqmuon}).
\bea
  \lim_{|\bq|\rightarrow 0_+} \langle \bq,e|L_\tau(t)|\bq,e \rangle
  \simeq -4 (s_{12} c_{12} s_{23})^2  \begin{cases}
    \cos(m_+ t)  \sin^2(\frac{m_-}{2}t)
      \qquad \text{when } \alpha_{21}=0, \\
    \cos(m_- t) \sin^2(\frac{m_+}{2}t)
      \qquad \text{when } \alpha_{21}=\pi.
    \end{cases}
 \label{eq:approxforsmallqtauv}
\eea
Then we note that mass squared differences are the same for both hierarchies,
\bea
m^2_2-m^2_1=m_+ m_-. 
\label{eq:product}
\eea
With the relation above, and Eqs.(\ref{eq:massnormal}-\ref{eq:massinverted}) one has,
\bea
m_+|_{\rm inverted} &\simeq& 5 \times m_+|_{\rm normal} \nn \\
m_-|_{\rm inverted} &\simeq& \frac{1}{5}  \times m_-|_{\rm normal}.
\eea
Therefore in the inverted hierarchical case, the longer period increases as $\frac{2 \pi}{m_-|_{\rm inverted}}\times 0.02 \sim 200$. (See the right figure of Fig.\ref{Fig5}.)
The shorter period decreases as $\frac{2\pi}{m_+|_{\rm inverted}} \times 0.02 \sim 1.1 $.
The latter point is shown in Fig.\ref{Fig6} by magnifying the scale of horizontal axis of
the right figure of Fig.\ref{Fig5}. The behavior of two curves of Fig.\ref{Fig6} is explained
with Eq.(\ref{eq:approxforsmallqtauv}) and the time $t=50(1/{\rm eV}) \tau$.   
The black curve shows the strongly suppressed
amplitude by $-\cos(m_+  t)  \sin^2(\frac{m_-}{2}t)$ within the time range 
satisfying $ \frac{m_-}{2}t \ll 1$.  While the red line shows the unsuppressed
amplitude by $-\cos(m_-  t)  \sin^2(\frac{m_+}{2}t)\simeq - \sin^2(\frac{m_+}{2}t)$.
Both curves have a period roughly given by  
$\frac{2\pi}{m_+|_{\rm inverted}}$.
\\
In  Fig.\ref{Fig7}, we have also plotted the momentum ($q$) dependence of the expectation value 
of the  electron number $\langle e(q)|L_e(\tau)|e(q) \rangle$ 
at  fixed dimensionless time $\tau$.
When the momentum is larger than the neutrino masses, the electron number stays in the range of [0,1] and it shows behavior similar to the energy dependence of the survival probability $P_{\rm{ee
}}$.
As the momentum decreases, the electron number oscillates and changes its sign. This indicates the transition to anti-neutrino through Majorana mass term. 
As the momentum is further reduced, the electron number reaches to a fixed value. 
\newpage
\section{Conclusions}
We formulate the time evolution of lepton family numbers under the presence of a Majorana mass term. 
The Majorana mass term is switched on at time $t=0$ and the lepton family number operator for arbitrary time is derived. 
Since the operator in the flavor eigenstate is continuously connected to that of the mass eigenstate, 
the creation and annihilation operators for the two eigenstates are related to each other. 
We also study the expressions for time evolutions of its expectation values. 

In our numerical calculations we show the time evolution of electron number, muon number, and tauon number for an electron neutrino with lepton number equal to one as the initial state.  We consider three cases of the neutrino's momentum when it is larger, as comparable as, or smaller than the assumed lightest neutrino mass ($m_1$ or $m_3$).
For the first case, we find that the lepton family numbers lie in the range from $0$ to $1$.
Next, we also show the case with the momentum which is comparable with the lightest neutrino mass.
We find that the lepton family number for the specific choice of Majorana phases can take negative values.
However most of the time, they are positive. We also find that the dependence on the Majorana phases is visible.
Finally, the case of the momentum which is smaller than that of the neutrino mass is shown.
We find that the lepton family number can take negative for arbitrary Majorana phases and we can see the differences caused by the choice of Majorana phases.

We comment on the implication for the study of C$\nu$B. In the early universe, the neutrinos decouple from the weak interaction, they are relativistic and the lepton family numbers are approximately conserved. As the universe expands, the neutrinos’ momenta decrease because of the redshift effect. At some stage, the neutrinos momenta are comparable to their rest mass. This situation could be mimicked by switching on the Majorana mass term with the step function. As time passes, eventually they become non-relativistic. In contrast to the above history of the C$\nu$B, we have considered the time evolution of lepton family numbers for neutrinos with a fixed momentum.

In future work, it would be interesting to study, in the Schrodinger picture, the time evolution of a state from a neutrino with a definite family number.  We also plan to separately make a comparison to neutrinos with Dirac mass.

\vspace{1cm}
\noindent
{\bf Acknowledgement}

The work of T.M. is supported by Japan Society for the Promotion of Science (JSPS) KAKENHI Grant Number JP17K05418 and
the work of Y.S. is supported  by Fujyukai Foundation.
We thank Mikihiko Nakao and Uma Sankar for useful comments.
We also thank the organizers and the participants of Pacific 2019.

\newpage
\vspace{1cm}
\noindent

\appendix
\section{The derivation of Eq.(\ref{eq:operators}) and Eq.(\ref{eq:operators2}).} \label{sec:appendixderivation}
In this appendix, we show the derivation of Eq.(\ref{eq:operators}) and Eq.(\ref{eq:operators2}), and the relationship of the momentum regions $A$ and $\bar{A}$ discussed in section \ref{ssec:oneflavor}.
We write the continuity condition of Eq.(\ref{eq:con1}) using the mode expansions for the Majorana field of Eq.(\ref{eq:massive}) and the massless field of Eq.(\ref{eq:massless}) at $t=0$. 
\bea
&&P_L \int^\prime \frac{d^3{\bf \bp}}{(2\pi)^3 2E(\bp)}\sum_{\lambda={\pm 1}}(a_M(\bp,\lambda)  u(\bp, \lambda)e^{i \bp \cdot \bx}+a_M^\dagger(\bp, \lambda)  v(\bp,\lambda)e^{-i \bp \cdot \bx}) \nn \\
&&=\int^\prime \frac{d^3 {\bf p}}{(2\pi )^32|\bp|} \left(a(\bp) e^{i \bp \cdot \bx} u_L(\bp)+b^\dagger(\bp) e^{-i \bp \cdot \bx}v_L(\bp) \right),
\label{eq:conappendix}
\eea
where, $P_L$ is the left-handed projection $P_L =\frac{1-\gamma_5}{2}$.  To derive the relation between operators of the Majorana field and that of the massless field, we split the non-zero momentum $\bp$ into two regions. 
For this purpose, we write the direction of the 
momentum $\bp$ with a
polar angle $\theta$ and an azimuthal angle $\phi$ as, 
\bea
\bn = \frac{\bp}{|\bp|}=\begin{pmatrix} \sin \theta \cos \phi \\ \sin \theta \sin \phi \\ \cos \theta \end{pmatrix}. 
\label{eq:angles}
\eea
Then, the non-zero momentum $\bp$  can be classified into two hemisphere regions defined as,
\bea
A=\{\bp  \in A| 0 \le \theta \le \pi, 0 \le \phi < \pi \} , \quad
\overline{A}=\{\bp  \in \overline{A}| 0 \le \theta \le \pi, \pi \le \phi < 2 \pi \}.
\label{eq:A}
\eea
We note that the change of the angles in Eq.(\ref{eq:angles}) from $(\theta, \phi) $ to $(\pi-\theta, \phi+\pi)$ results in a one to one
 mapping from region $A$ to region $\bar{A}$.  Also, this results in a change to the sign of $\bn$.
Using the one to one mapping, one can rewrite the continuity condition of Eq.(\ref{eq:conappendix}) with the momentum integration only over the region $A$.
\begin{multline}
P_L \int_{\bp \in A} \frac{d^3{\bf \bp}}{(2\pi)^3 2E(\bp)}\sum_{\lambda={\pm 1}} \{
(a_M(\bp,\lambda)  u(\bp, \lambda)+a_M^\dagger(-\bp,\lambda) v(-\bp,\lambda) )e^{i  \bp \cdot \bx } \\
\shoveright{ + (a_M^\dagger(\bp, \lambda)  v(\bp,\lambda)+a_M(-\bp, \lambda)  u(-\bp,\lambda)) e^{-i \bp \cdot \bx}  \}} \\
=\int_{\bp \in A} \frac{d^3 {\bf p}}{(2\pi )^32|\bp|}  \{ (a(\bp) u_L(\bp)+b^\dagger(-\bp) v_L(-\bp) ) e^{i \bp \cdot \bx} \\
+ (a(-\bp) u_L(-\bp)+b^\dagger(\bp) v_L(\bp))   e^{-i \bp \cdot \bx} \}.
\label{eq:cc}
\end{multline}
Eq.(\ref{eq:cc}) is translated into the following relations,
\begin{gather}
\frac{1}{2E(\bp)}\sum_{\lambda={\pm 1}} (a_M(\bp,\lambda) P_L u(\bp, \lambda)+a_M^\dagger(-\bp,\lambda) P_L v(-\bp,\lambda) )=\frac{1}{2|\bp|} (a(\bp) u_L(\bp)+b^\dagger(-\bp) v_L(-\bp) ), \\
\frac{1}{2E(\bp)}\sum_{\lambda={\pm 1}} (a_M^\dagger(\bp, \lambda) P_L v(\bp,\lambda)+a_M(-\bp, \lambda) P_L u(-\bp,\lambda)) = \frac{1}{2|\bp|} (a(-\bp) u_L(-\bp)+b^\dagger(\bp) v_L(\bp)). 
\label{eq:con2}
\end{gather}
The left-handed $P_L=\frac{1-\gamma_5}{2}$ projection of the spinors are obtained 
from  Eq.(\ref{eq:massspinorcon1}) and Eq.(\ref{eq:massspinorcon2}) as,
\bea
P_L u(\pm\bp, +1)  &=&\sqrt{N(\bp)}
        \begin{pmatrix} 0 \\  \frac{m}{E(\bp)+\ap} \phi_+(\pm\bn) \end{pmatrix},\quad P_L u(\pm\bp, -1)=\sqrt{N(\bp)}
        \begin{pmatrix} 0  \\ \phi_-(\pm\bn) \end{pmatrix},  \nn \\
P_L v(\pm\bp, +1)&=&\sqrt{N(\bp)}
        \begin{pmatrix} 0 \\ -\phi_-(\pm\bn) \end{pmatrix}, \quad P_L v(\pm\bp,-1)= \sqrt{N(\bp)}
        \begin{pmatrix}0  \\  \frac{m}{E(\bp)+\ap} \phi_+(\pm\bn) \end{pmatrix}, \nn \\
\label{eq:projected}
\eea
where the two component spinors $\phi_\pm(\pm\bn)$ are written by angles $\theta$ and $\phi$ from Eq.(\ref{eq:angles}) as,
\bea
\phi_+(\bn)&=&\begin{pmatrix} e^{-i \frac{\phi}{2}} \cos\frac{\theta}{2} \\ e^{i \frac{\phi}{2}} \sin\frac{\theta}{2} \end{pmatrix}, \quad
\phi_-(\bn)= \begin{pmatrix} - e^{-i \frac{\phi}{2}} \sin\frac{\theta}{2} \\ e^{i \frac{\phi}{2}} \cos\frac{\theta}{2} \end{pmatrix}, \nn \\
\phi_+(-\bn)&=&\begin{pmatrix} -i e^{-i \frac{\phi}{2}} \sin\frac{\theta}{2} \\  i e^{i \frac{\phi}{2}} \cos\frac{\theta}{2} \end{pmatrix} =i \phi_-(\bn), \quad
\phi_-(-\bn)= \begin{pmatrix} i e^{-i \frac{\phi}{2}} \cos\frac{\theta}{2} \\ i e^{i \frac{\phi}{2} } \sin \frac{\theta}{2} \end{pmatrix}=i \phi_+(\bn).
\label{eq:two}
\eea
Using Eq.(\ref{eq:con2}), Eq.(\ref{eq:projected}), and Eq.(\ref{eq:two}) one obtains 
the operator relations,
\bea
\frac{a(\bp)}{\sqrt{2|\bp|}}&=& \frac{\sqrt{N(\bp)}}{2E(\bp)} (a_M(\bp,-)+i \frac{m}{E(\bp)+|\bp|} a^\dagger_M(-\bp,-) ) ,\\
\frac{b^\dagger(-\bp)}{\sqrt{2|\bp|}}&=& \frac{\sqrt{N(\bp)}}{2E(\bp)} (a^\dagger_M(-\bp,+)+i \frac{m}{E(\bp)+|\bp|} a_M(\bp,+) ),\\
\frac{a(-\bp)}{\sqrt{2 |\bp|}}&=& \frac{\sqrt{N(\bp)}}{2E(\bp)} (a_M(-\bp,-)-i \frac{m}{
E(\bp)+|\bp|} a^\dagger_M(\bp,-)), \\
\frac{b^\dagger(\bp)}{\sqrt{2 |\bp|}}&=& \frac{\sqrt{N(\bp)}}{2E(\bp)} (a^\dagger_M(\bp,+)-i \frac{m}{E(\bp)+|\bp|} a_M(-\bp,+) ),
\eea
where $\bp \in A$. This completes the derivation of Eq.(\ref{eq:operators}) and Eq.(\ref{eq:operators2}).

\end{document}